\documentclass[11pt,a4paper]{article}
\usepackage[T1]{fontenc}

\usepackage{jheppub}
\usepackage{mathtools}
\usepackage{physics}
\usepackage{tikz-cd}
\usepackage{mathrsfs}
\usepackage{slashed}

\usepackage{natbib}
\newcommand{\pa}{\mathop{}\!\partial}

\usepackage{mathtools}

\newcommand{\pth}{\text{\normalfont\TH\/}}

\newcommand{\R}{\mathbb{R}}

\def\be#1\ee{\begin{align}#1\end{align}}

\def\heq{\,\hat{=}\,}

\newcommand\nueq{\stackrel{\mathclap{\normalfont\tiny\mbox{$\cH$}}}{=}}

\newcommand{\diff}{\mathfrak{diff}}

\newcommand{\mr}{\mathring}
\newcommand{\fr}{\mathfrak}
\newcommand{\bd}{\boldsymbol}

\newcommand{\dext}{\text{d}}

\newcommand{\cS}{\mathcal{S}}

\newcommand{\cN}{\mathcal{N}}

\newcommand{\cA}{\mathcal{A}}
\newcommand{\cF}{\mathcal{F}}

\newcommand{\cO}{\mathcal{O}}
\newcommand{\Lie}{\mathcal{L}}

\newcommand{\cE}{\mathcal{E}}

\newcommand{\cH}{\mathcal{H}}

\newcommand{\cQ}{\mathcal{Q}}
\newcommand{\cT}{\mathcal{T}}
\newcommand{\cP}{\mathcal{P}}
\newcommand{\cJ}{\mathcal{J}}

\newcommand{\cR}{\mathcal{R}}

\newcommand{\scr}{\mathscr}

\newcommand{\eps}{{\bd{\epsilon}}}

\newcommand{\f}{\frac}
\newcommand{\wt}{\widetilde}

\def\EC{\mathrm{EC}}
\def\ECH{\mathrm{ECH}}
\def\H{\mathrm{H}}
\def\TF{\mathrm{TF}}

\def\scri{\mathscr{I}}

\newcommand{\mtn}{\mr{\theta}^{(n)}}
\newcommand{\mpi}{\mr{\pi}}

\newcommand{\thell}{\theta^{(\ell)}}
\newcommand{\mthell}{\mr{\theta}^{(\ell)}}
\newcommand{\mkell}{\mr{\kappa}_{(\ell)}}
\newcommand{\sigell}{\sigma^{(\ell)}}

\newcommand{\loplus}{%
\begin{tikzpicture}%
\hspace{1.5pt} \draw (0,0) circle (3.5pt);
\draw (0,3.5pt) -- (0,-3.5pt);
\draw (0,0) -- (3.5pt,0); 
\end{tikzpicture}
\hspace{3pt}
}

\allowdisplaybreaks

\title{Duality symmetry and dynamics on finite null boundaries}
\author[a,b]{Gianfranco De Simone,}
\affiliation[a]{Università degli Studi di Udine,\\ via Palladio 8, I-33100 Udine, Italy}
\affiliation[b]{National Institute for Nuclear Physics (INFN), Sezione di Trieste,\\ Via Valerio 2, 34127, Italy}
\emailAdd{gianfranco.desimone@uniud.it}

\abstract{ In this work, we derive a set of boost-weighted $w$  functionals of the metric, with $w\in\{2,1,0,-1,-2\}$,  which transform semi-covariantly under the action of the near-horizon symmetry group.
In particular, we demonstrate that the knowledge of the $w=-2$ metric functional and its behaviour under the near-horizon symmetry transformations allows us to derive the expressions and the properties of the remaining boost-weighted functionals via a recursive pattern. A similar recursive pattern also appears when evaluating the action of the near-horizon symmetry group on the evolution equations of these boost-weighted functionals. Again, the knowledge of the evolution equation of the boost $w=-2$ functional and its behaviour under symmetry transformations allows the remaining evolution equations to be determined using symmetry arguments. We also emphasize the role played by the duality symmetry in the characterization of the phase space of a null boundary and in the evaluation of the equation of motion. In conclusion, we derive the sub-leading Noether charges in the Einstein-Cartan-Holst formulation of gravity, showing that the imaginary part of the Weyl scalar $\Psi_2$ appears at the sub-leading order in the Holst charge.}

\begin{document}

\maketitle
\flushbottom

\section*{Introduction} \addcontentsline{toc}{section}{Introduction}
The analysis of the geometry of null hyper-surfaces has contributed significantly to advancements in gravitational physics. Over the past decade, the study of the symmetries and the construction of the phase space on null boundaries have yielded a wealth of new results, paving the way toward a holographic description of gravity. Specifically, two primary classes of null boundaries are distinguished: the asymptotic null boundaries (such as the (past) future null infinity ($\scri^-$) $\scri^+$ in asymptotically flat spacetimes) and  the null boundaries located at a finite distance (such as the (past) future horizon ($\scr{H}^-$) $\scr{H}^+$ of a black hole). Efforts have been made to relate the physics of finite null hyper-surfaces with that of asymptotic null boundaries.

\vspace{0.2cm}

\noindent
The analysis of asymptotic symmetries dates back to the early sixties, when Bondi, van der Burg, Metzner and Sachs demonstrated that the gravitational symmetry group in a radiative, asymptotically flat spacetime consists of the semi-direct product of the Lorentz group with super-translations (an infinite-dimensional Abelian sub-group), forming the so-called BMS group \cite{Bondi:1960jsa,Bondi:1962px, Sachs:1961zz, Sachs:1962wk, Madler:2016xju}. However, it has only been established in the last decade that this symmetry group (and its various extensions studied in \cite{Barnich:2009se, Barnich:2010eb, Barnich:2011mi, Campiglia:2014yka, Campiglia:2015yka, Compere:2018ylh, Freidel:2021fxf, Geiller:2022vto, Geiller:2024amx}) plays a fundamental role in understanding the infrared structure of gravitational scattering \cite{Strominger:2017zoo}. In particular, this infrared structure can be encapsulated in the so-called (leading) infrared triangle, which ties together super-translations, the displacement memory effect \cite{Braginsky:1987kwo,Christodoulou:1991cr}, and the Weinberg soft graviton theorem \cite{Weinberg:1965nx}. Nonetheless, these relations represent only the tip of the iceberg, as a richer structure emerges when considering the subleading contributions in the asymptotic expansion. Indeed, at subleading order a new infrared picture appears, connecting super-Lorentz transformations \cite{Campiglia:2014yka}, the spin memory effect \cite{Pasterski:2015tva}, and the subleading soft graviton theorem \cite{Cachazo:2014fwa}; going deeper, the sub-subleading soft graviton theorem dictates the conservation of a spin-2 charge generating a non-local symmetry \cite{Freidel:2021dfs}. Ultimately, exploring the sub-leading structure of asymptotically flat spacetimes in further depth, an infinite tower of soft symmetries emerges packaged in the $w_{1+\infty}$ algebra, first discovered in \cite{Strominger:2021mtt} and subsequently investigated in \cite{Freidel:2021ytz, Geiller:2024bgf} through a canonical and a non-perturbative analysis \cite{Cresto:2024fhd, Cresto:2024mne} and also in  \cite{Adamo:2021lrv, Donnay:2024qwq, Kmec:2024nmu} from a twistor perspective.

\vspace{0.2cm}

\noindent
On the other hand, significant progress has been achieved regarding null boundaries located at a finite distance \cite{Donnay:2015abr, Donnay:2016ejv, Chandrasekaran:2018aop, Donnay:2019jiz, Ashtekar:2021wld, Ashtekar:2021kqj, Liu:2022uox, Ashtekar:2024bpi, Ashtekar:2024stm}. BMS-like symmetries have been explored for arbitrary null hyper-surfaces, leading to remarkable results and providing new insights into the black hole information loss problem \cite{Hawking:2016sgy, Hawking:2016msc, Haco:2018ske}.
Furthermore, intriguing results have been established between Carrollian physics and stretched horizons, providing novel links to a hydrodynamical description of black holes \cite{Penna:2015gza, Penna:2018gfx, Donnay:2019jiz, Freidel:2022vjq, Freidel:2024emv}.\\
Nonetheless, relevant differences occur between asymptotic and finite distance null boundaries, most notably the presence of genuine degrees of freedom in the latter case. Indeed, asymptotic null boundaries behave as weakly isolated horizons \cite{Ashtekar:2021kqj, Ashtekar:2021wld, Ashtekar:2024bpi, Ashtekar:2024stm, Agrawal:2025fsv} and these degrees of freedom are frozen due to the boundary conditions. Therefore, a direct correspondence has been identified between weakly isolated horizons and null infinity \cite{Ciambelli:2025mex, Agrawal:2025fsv}.\\
However, it has recently been claimed in \cite{Ruzziconi:2025fct, Ruzziconi:2025fuy} that an asymptotic-finite correspondence should be investigated by analysing the sub-leading phase space of finite distance null hyper-surfaces, where the self duality condition must be imposed. This self-duality condition allows for the identification of an analogue of the previously discussed asymptotic celestial $w_{1+\infty}$ symmetry. 

\vspace{0.2cm}

\noindent
In this work, we continue the investigation of symmetries on a finite distance null hyper-surfaces, emphasizing similarities and differences with null infinity. The main goal of the manuscript is to construct functionals of the metric which transform semi-covariantly under the action of the near-horizon symmetry group and to subsequently identify their evolution equations via symmetry arguments. In the same spirit of \cite{Freidel:2021qpz}, we identify in the sub-sub-leading order of the space-like boundary metric a boost-weighted $w=-2$ quantity that transforms
covariantly under boost transformations and semi-covariantly under the near-horizon symmetry group. This boost-($-2$) tensor, denoted by $\cT_{ab}$, corresponds to the leading order of the Weyl scalar $\Psi_4$ and represents the spin $s=-2$ charge when contracted with the complex dyad ($\mr{E}^a, \mr{\bar{E}}^a$), as shown in \cite{Ruzziconi:2025fuy}. Once the behaviour  of the covariant functional $\cT_{ab}$ under the action of the symmetry transformations is established, the expressions of the remaining boost covariant functionals follow straightforwardly by observing that the following rule
\begin{equation}
    \var_{(\tau, Y)}\mathrm{Q}_{w} = (\tau\pa_v + \Lie_Y +w\dot{\tau}) \mathrm{Q}_w -(w-2)\mathrm{Q}_{w+1}\pa\tau
\end{equation}
holds, for $\mathrm{Q}_w \in\{\cN_{ab},\cJ_{a}, \cA_{ab}, \cP_a, \cT_{ab}\}$ and $w\in\{2,1,0,-1,-2\}$. Moreover, by noticing that the time derivative acts as a boost-raising operator, we then look for combinations of these metric functionals that transform homogeneously under the near-horizon symmetry group. Therefore, using the same strategy adopted in \cite{Freidel:2021qpz} for null infinity, we identify the following combinations:
\begin{equation}
\begin{aligned}
& \pa_v \cJ_a -\Bigl(\mkell-\f{3}{2}\mthell\Bigl)\cJ_a - \mr{\sigma}^{(\ell)b}_{a} \cJ_b- (\mr{D}_b -\mpi_b)\cN^{\ b}_a=0,\\
&\pa_v\cA +\f{3}{2}\mthell\cA - (\mr{D}_a + \mpi_a)\cJ^a + \mr{\sigma}^{(n)}_{ab}\cN^{ab} =0,\\
&\pa_v\wt{\cA} +\f{3}{2}\mthell \wt{\cA} - (\mr{D}_a + \mpi_a)\wt{\cJ}^a + \mr{\sigma}^{(n)}_{ab}\wt{\cN}^{ab} =0,\\
&\pa_v\cP_a + \mr{\mu}_{(\ell)} \cP_a -\mr{\sigma}^{(\ell)b}_{a}\cP_b  -\f{1}{2}(\pa_a +3\mpi_a)\cA - \f{1}{2}(\wt{\pa}_a +3\mr{\wt{\pi}}_a)\wt{\cA} +2\mr{\sigma}^{(n)}_{ab}\cJ^b =0,\\
&\pa_v\cT_{ab} +\Bigl(2\mkell -\f{1}{2}\mthell \Bigl) \cT_{ab}  - (\mr{D}_{(a}+5\mpi_{(a}) \cP_{b)} +\f{3}{2}\mr{\sigma}^{(n)}_{ab}\cA +\f{3}{2}\mr{\wt{\sigma}}^{(n)}_{ab}\wt{\cA} =0,
\end{aligned}
\label{0.5}
\end{equation}
whose behaviour under near-horizon symmetry transformations can be summarized through the following recursive pattern
\begin{equation}
    \cE^{\mathrm{Q}_w} = (\tau\pa_v + \Lie_Y +(w+1)\dot{\tau}) \cE^{\mathrm{Q}_w} +(2-w)\cE^{\mathrm{Q}_{w+1}}\pa\tau.\label{EQS}
\end{equation}
Hence, once the evolution equation of the lowest boost-weighted covariant functional is established, the evolution equations of the others follow immediately from \eqref{EQS}. In the set of equations shown above, the tilde denotes the duality operation discussed in subsection \ref{area_dual}. Similar to the null infinity analysis \cite{Freidel:2021qpz}, the duality symmetry plays a fundamental role in evaluating the dynamics of the covariant functionals and encapsulates the concept of the dual gravitational charges \cite{Godazgar:2018dvh, Godazgar:2018qpq, Godazgar:2020kqd, Oliveri:2020xls}. Furthermore, the structure of \eqref{EQS} is significantly richer than its asymptotic analogue, reflecting the fact that null infinity behaves specifically as a weakly isolated horizon \cite{Ashtekar:2021kqj, Ashtekar:2021wld, Ashtekar:2024bpi, Ashtekar:2024stm, Agrawal:2025fsv}.

\vspace{0.2cm}

\noindent
Finally, as demonstrated in \cite{Ruzziconi:2025fuy}, the evaluation of the sub-leading near-horizon phase space reveals a structure analogous to the leading phase space at null infinity. Indeed, the authors in \cite{Ruzziconi:2025fuy} showed the sub-leading Barnich-Brandt charges are related to the leading orders of the Weyl scalars $\Psi_3$ and $\Re\{ \Psi_2\}$. Adding a topological term to the Einstein-Cartan Lagrangian, we demonstrate that the imaginary part of the Weyl scalar $\Psi_2$ also enters into the sub-leading super-translation Noether charge, emphasizing the similarities with asymptotic dual gravitational charges identified in \cite{Godazgar:2018dvh, Godazgar:2018qpq, Godazgar:2020kqd}.

\vspace{0.2cm}

\noindent
The organization of the paper is as follows. The first section collects some results concerning the geometry near a null hyper-surface. Some of the results follow from \cite{DeSimone:2025ouu} and reference therein, employing the metric ansatz introduced in \cite{Booth:2006bn, Booth:2012xm}.\\
The second section encapsulates the core of the work. Following the analysis carried out in \cite{Freidel:2021qpz} for asymptotic null boundaries, we construct operators of boost weights $w=-2,-1,0,1,2$ which transform homogeneously under the homogeneous subgroup of the near-horizon symmetry group. In particular, we emphasize the role of duality transformation, which is instrumental in arranging the Bianchi identities in appendix \ref{NP_an} into a suitable metric form. \\
The third section is devoted to analysing the behaviour of the equations of motion of the covariant functionals under the near-horizon symmetry group. We find that the same pattern found at null infinity also holds at a finite distance.\\
Finally, in section \ref{sec4} we compute the sub-leading Einstein-Cartan-Holst Noether charges by using the covariant phase space methods, and compare our charges with those obtained in \cite{Ruzziconi:2025fuy}. Although the sub-leading charge associated with $\cS$-diffeomorphisms coincides with the result in \cite{Ruzziconi:2025fuy}, the sub-leading charge relative to super-translations does not. A similar discrepancy was encountered at null infinity in \cite{Freidel:2021fxf} and resolved through the introduction of a non-covariant boundary Lagrangian. This remarkable result achieved in \cite{Freidel:2021fxf} highlights the importance of the Noetherian construction advocated in \cite{Freidel:2021cjp}, alongside the definition of a Lagrangian-dependent charge bracket. Consequently, following the same strategy, we conclude by providing the expression for the non-covariant boundary Lagrangian that yields the Noetherian construction of the sub-leading charges obtained in \cite{Ruzziconi:2025fuy}.\\
For completeness, we have set up the calculation of Einstein-Cartan-Holst Noetherian fluxes in appendix \ref{app_A} and provided a detailed Newman-Penrose analysis in appendix \ref{NP_an}.

\vspace{0.5cm}

\noindent
\emph{Notations and conventions}: In this work, natural units and the metric signature $(-+++)$ are adopted. Greek letters $\mu,\nu,...$ denote spacetime indices, while lowercase Latin letters $a,b,\cdots$ are used for indices on the corner $\cS$. Uppercase Latin letters are used for spinorial indices. In particular, lowercase Latin letters from $i,j,\cdots$, are used to denote the complex dyad, i.e. $e^i_\mu=(m_\mu, \bar{m}_\mu)$. The value of a quantity evaluated on the horizon $\cH$, namely when $\rho=0$, is denoted by a circle over such a quantity, i.e. $\mthell \nueq \thell$. To symmetrize (resp. anti-symmetrize) indices we use the round (resp. square) brackets $(\cdot)$ (resp. $[\cdot]$) and we denote by $\langle \cdot \rangle$ the symmetric trace free part. The volume form is $\eps = \sqrt{-g}\ \dext^4 x$ with orientation $\epsilon_{v\rho\theta\phi}=1$. We follow the notation used in \cite{Freidel:2021cjp} for the covariant phase space formalism.

\section{Near-horizon geometry: review}\label{review1}
To begin with, in this section we recall some results concerning null geometry and near-horizon symmetries \cite{Booth:2006bn, Booth:2012xm, Chandrasekaran:2021hxc, Chandrasekaran:2023vzb,DeSimone:2025ouu}. Let us introduce the Gaussian null coordinates $(v, \rho, x^a)$, where $v$ is the advanced time, $\rho$ is the radial coordinate and $x^a$ denotes the coordinates on the space-like codimension-2 surface $\cS$. The null boundary (/horizon) $\cH$ is defined as a smooth union of the space-like hyper-surfaces $\cS_v$, namely $\cH = \cup_v \cS_v$. In these coordinates, the near-horizon metric assumes the following form 
\begin{equation}
    \dext s^2 = -2\dext v\dext \rho + 2 V\dext v^2 +q_{ab}(\dext x^a+ U^a\dext v)(\dext x^b + U^b \dext v),
    \label{metric}
\end{equation}
which satisfies the so-called Newman-Unti \cite{Newman:1962cia} (NU) gauge conditions
\begin{equation}
    g_{\rho\rho}=0, \qquad g_{\rho a}=0, \qquad g_{v\rho}=-1,
\label{NU_gc}
\end{equation}
and the inverse metric is 
\begin{equation}
    g^{\mu\nu}\pa_\mu\pa_\nu = -2\pa_v\pa_\rho -2V\pa_\rho\pa_\rho +2U^a\pa_a\pa_\rho +q^{ab}\pa_a\pa_b.
\end{equation}
The Levi-Civita connection compatible with the boundary metric $q_{ab}$ is denoted by $D_a$, and by $\cR_{ab}[q]$ and $\cR[q]$ we identify the relative Ricci tensor and Ricci scalar. In addition to the gauge conditions imposed above, we require the metric to satisfy the following boundary conditions,
\begin{equation}
    g_{vv}=O(\rho), \qquad g_{va}=O(\rho), \qquad g_{ab} = O(1).\label{bc}
\end{equation}
Now, in order to extract some relevant physical quantities for our analysis,  let us introduce two null vectors
\begin{equation}
{\ell}^\mu \pa_\mu =  \pa_v +V\pa_\rho- U^a \pa_a\qquad \text{and} \qquad {n}^\mu= \pa_\rho,\label{l_e_n}
\end{equation}
which span the space normal to $\cS$. These two null vectors are defined up to a local rescaling
\begin{equation}
{\ell}\to e^{\lambda_L}{\ell} \qquad \text{and}\qquad {n}\to e^{-\lambda_L}{n},
\label{boost_res}
\end{equation}
where $\lambda_L(v, \sigma^c)$ is a smooth function on the null boundary $\cH$. The null vectors \eqref{l_e_n} define the following extrinsic curvatures
\begin{equation}
    K^{(\ell)}_{ab} = \f{1}{2}\Bigl(\pa_v q_{ab} +V\pa_\rho q_{ab} - 2D_{(a}U_{b)}\Bigl), \qquad\text{and}\qquad
    K^{(n)}_{ab} = \f{1}{2} \pa_\rho q_{ab},
\end{equation}
where the trace-free part of $K^{(\ell)}_{ab}$ (resp. $K^{(n)}_{ab}$) defines the longitudinal (resp. transversal) shear $\sigma^{(\ell)}_{ab}$ (resp. $\sigma^{(n)}_{ab}$), while its trace defines the longitudinal (resp. transversal) expansion $\thell$ (resp. $\theta^{(n)}$). Under the rescaling \eqref{boost_res}, the extrinsic curvatures transform as follows
\begin{equation}
K^{(\ell)}_{ab}\to e^{\lambda_L}K^{(\ell)}_{ab}, \qquad K^{(n)}_{ab}\to e^{-\lambda_L}K^{(n)}_{ab}.
\end{equation}
Moreover, we introduce the rotational $1-$form
\begin{equation}
    \varpi_\mu = -  n_\nu\nabla_\mu \ell^\nu \label{rot-1f},
\end{equation}
whose spatial projection, denoted by $\pi_a$, defines the Hájiček field. In particular, the contraction $\ell^\mu \varpi_\mu$ is equal to the non-affinity parameter $\kappa_{(\ell)}$, defined by
\begin{equation}
\ell^\mu\nabla_\mu \ell^\nu = \kappa_{(\ell)} \ell^\nu.
\end{equation}
The Hájiček field and the non-affinity parameter transform as connections under \eqref{boost_res}, i.e.,
\begin{equation}
    \pi_a \to \pi_a + \pa_a\lambda_L,\qquad \kappa_{(\ell)} \to e^{\lambda_L}(\kappa_{(\ell)} + \ell^\mu\pa_\mu\lambda_L).
\end{equation}
Once the non-affinity parameter and the Hájiček field are evaluated, the explicit expressions of metric function $V(x^\mu)$ and the vector $U^a(x^\mu)$ in \eqref{metric} are derived from the following relations
\begin{equation}
    V = \int \dext\rho\ \kappa_{(\ell)}, \qquad \text{and}\qquad U^a =2 \int \dext\rho\ \pi^a .\label{FU}
\end{equation}
The radial behaviour of these functions is encoded in the hypersurface Einstein equations, evaluated in the next subsection.

\subsection{Einstein equations}
In order to solve the hypersurface Einstein equations
\begin{equation}
    \mathbb{E}_{\rho\mu} := R_{\rho\mu}-\f{1}{2}Rg_{\rho\mu} +\Lambda g_{\rho\mu} =0,
    \label{Eins_eq}
\end{equation}
we assume the following radial behaviour of the boundary metric
\begin{equation}
q_{ab}(\rho, x^i)=\mr{q}_{ab}(x^i) + \rho\lambda_{ab}(x^i) + \rho^2 d_{ab}(x^i) + \rho^3 d^{(1)}_{ab} (x^i) + O(\rho^4),
\label{q_metric_exp}
\end{equation}
denoting by $\mr{D}_a$ the covariant derivative of the leading order metric $\mr{q}_{ab}$, i.e.,
\begin{equation}
D_a U^b = \mr{D}_a U^b + O(\rho). 
\end{equation}
Thus, solving the hypersurface Einstein equations, we obtain the following radial behaviours
\begin{equation}
\begin{aligned}
\theta^{(n)}(x^\mu) &= \mtn - \rho \mr{K}^{(n)}_{cd}\mr{K}^{cd}_{(n)} +o(\rho),\\
\pi_a(x^\mu) &= \mr{\pi}_a (v,x^c) + \rho\Bigl( \mr{D}_b\mr{K}_a^{(n)b} - (\mr{D}_a +\mr{\pi}_a)\mr{\theta}^{(n)}\Bigl) +o(\rho),\\
\kappa_{(\ell)}(x^\mu) &= \mr{\kappa}_{(\ell)}(v, x^c) + \rho \Bigl(-\mr{K}^{ab}_{(n)} \mr{K}^{(\ell)}_{ab} - (\pa_v + \mr{\kappa}_{(\ell)})\mr{\theta}^{(n)} + (\mr{D}_a -2\mr{\pi}_a)\mr{\pi}^a\\
&\quad + \f{1}{2}(\mr{R}- 2\Lambda)  \Bigl) + o(\rho).
\end{aligned}
\end{equation}
From the relations in \eqref{FU}, we obtain the following radial expansions of the metric functions,
\begin{equation}
\begin{aligned}
V&= \rho\mr{\kappa}_{(\ell)} + \f{\rho^2}{2} \Bigl(-\mr{K}^{ab}_{(n)}\mr{K}^{(\ell)}_{ab} - (\pa_v + \mr{\kappa}_{(\ell)})\mr{\theta}^{(n)} + (\mr{D}_a -2\mr{\pi}_a)\mr{\pi}^a + \f{1}{2}(\mr{R}-2\Lambda)\Bigl) + o(\rho^2),\\
U^a&= 2\rho\mr{\pi}^a + \rho^2\Bigl(\mr{D}_b\mr{K}^{ab}_{(n)} - (\mr{D}^a + \mr{\pi}^a) \mr{\theta}^{(n)} - 2\mr{K}_{(n)}^{ab}\mr{\pi}_b\Bigl) + o(\rho^2),\\ 
q_{ab} &= \mr{q}_{ab} + 2\rho\mr{K}^{(n)}_{ab}+ \rho^2 \Bigl(d_{\langle ab \rangle} +\f{1}{2} \mr{q}_{ab}\mr{K}^{(n)}_{cd}\mr{K}^{cd}_{(n)}  
\Bigl) + o(\rho^2).
\end{aligned}
\end{equation}
Therefore, the metric up to the second order in the radial coordinate reads 
\begin{equation}
\begin{aligned}
    \dext s^2 &= -2\dext v\dext \rho + \mr{q}_{ab} \dext x^a \dext x^b \\
    &\quad +2\rho\Bigl\{ \mr{\kappa}\dext v^2 
    +2\mr{\pi}_a \dext x^a \dext v
    + \mr{K}^{(n)}_{ab}\dext x^a \dext x^b\Bigl\}\\
    & \quad + \rho^2 \Bigl\{\Bigl(-\mr{K}^{ab}_{(n)}\mr{K}^{(\ell)}_{ab} - (\pa_v + \mr{\kappa})\mr{\theta}^{(n)} + (\mr{D}_a +2\mr{\pi}_a)\mr{\pi}^a+ \f{1}{2}(\mr{R}-2\Lambda)\Bigl)  \dext v^2 \\
    &\quad  + 2\Bigl((\mr{D}_b { + 2\mpi_b})\mr{K}^{(n)b}_{a} - (\mr{D}_a + \mr{\pi}_a) \mr{\theta}^{(n)} \Bigl) \dext x^{a}\dext v +\Bigl( d_{\langle ab\rangle} + \f{1}{2}\mr{q}_{ab} \mr{K}_c^{(n)d} \mr{K}^{(n)c}_d\Bigl)\dext x^a \dext x^b \Bigl\}\\
    &\quad + O(\rho^3).
\end{aligned}
\end{equation}
The Einstein evolution equations can be computed by evaluating order by order in the radial coordinate the remaining components of the Einstein equations. The leading order of the null component $\mathbb{E}_{\ell\ell}$ yields
\begin{equation}
\begin{aligned}
\mr{\mathbb{E}}_{\ell\ell}=\pa_v \mr{\theta}^{(\ell)} -\mr{\kappa}_{(\ell)} \mr{\theta}^{(\ell)}  + \mr{K}_{(\ell)}^{ab}\mr{K}^{(\ell)}_{ab} ,
\end{aligned}
\label{null_ray}
\end{equation}
that is the null Raychaudhuri equation and the leading order of $\mathbb{E}_{\ell a}$ gives
\begin{equation}
\begin{aligned}
\mr{\mathbb{E}}_{\ell a}=(\pa_v +\mr{\theta}^{(\ell)})\mr{\pi}_a -\mr{D}_a (\mr{\kappa}_{(\ell)}+\mr{\theta}^{(\ell)}) +\mr{D}_b \mr{K}_{a}^{(\ell)b},
\end{aligned}
\label{damour}
\end{equation}
that is the Damour equation. The evolution equation of the transversal extrinsic curvature is encoded into the $\mathbb{E}_{ab}$ component, which reads
\begin{equation}
\begin{aligned}
 \mr{\mathbb{E}}_{ab} &= 2(\pa_v +\mr{\kappa}_{(\ell)})\mr{K}^{(n)}_{ab} -2\mr{D}_{(a}\mr{\pi}_{b)} -2\mr{\pi}_a\mr{\pi}_b + \mr{\theta}^{(n)} \mr{K}^{(\ell)}_{ab} + \mr{\theta}^{(\ell)} \mr{K}^{(n)}_{ab} \\
&\qquad - 2\mr{K}^{(\ell)}_{c(a}\mr{K}^{(n)c}_{b)} - 2\mr{K}^{(n)}_{c(a}\mr{K}^{(\ell)c}_{b)} + \mr{\cR}_{ab} -\f{1}{2}\mr{q}_{ab}(\mr{R}-2\Lambda),
\end{aligned}
    \label{Rab}
\end{equation}
where $\mr{\cR}_{ab}$ is the Ricci tensor associated with the boundary metric $\mr{q}_{ab}$. In particular, the trivial equation is
\begin{equation}
    \mr{\mathbb{E}}^c_{\ c} = \mr{\cR} + 2(\pa_v+\mr{\kappa}_{(\ell)})\mtn -2\mpi_c\mpi^c - 2\mr{D}_c\mpi^c + 2\mtn\mthell -\mr{R} + 2\Lambda.
\label{E^c_c}
\end{equation}
In the rest of the work, we set the cosmological constant to zero, since it does not play a fundamental role as in the asymptotic analysis.

\subsection{Near-horizon symmetries}\label{sec1.2}
In this subsection, we complete the review of the near-horizon geometry by introducing the spacetime diffeomorphisms that preserve the gauge and boundary conditions defined previously in \eqref{NU_gc} and \eqref{bc}. One finds that
\begin{equation}
\begin{aligned}
    \xi^v = \tau(v,x^c),\qquad
    \xi^a = Y^a(x^c) + I^{ab} \pa_b \tau,\qquad
    \xi^\rho = -\rho \dot{\tau} + I^{b}\pa_b\tau,
\end{aligned}
\end{equation}
where 
\begin{equation}
    I^{ab} = \int \dext\rho \ g^{ab} \qquad\text{and}\qquad I^b = \int \dext\rho\ g_{va}g^{ab}.
\end{equation}
Using the radial expansions furnished in the previous subsection, we have
\begin{equation}
\begin{aligned}
I^{ab} = \rho \mr{q}^{ab}- \rho^2\mr{K}_{(n)}^{ab} + o(\rho^2), \qquad 
I^b = \rho^2\mpi^b + o(\rho^2).   
\end{aligned}
\end{equation}
By truncating the $v$-expansion of $\tau$ at the linear level, i.e. $\tau = T(x^c) + vW(x^c)$, the algebra reads as a double semi-direct sum ,
\begin{equation}
     \fr{g} = (\diff(\cS)\loplus \R_W^{\cS})\loplus \R_T^{\cS},
\end{equation}
where the super-translations, Weyl super-boost and $\cS$-diffeomorphisms are generated by
\begin{equation}
\begin{aligned}
\xi_T &= T\pa_v + \rho \mr{q}^{ab}\pa_b T \pa_a + \rho^2 \pa_b T\Bigl( \mr{\pi}^b \pa_\rho - \mr{K}_{(n)}^{ab} \pa_a \Bigl)+O(\rho^3),\\
\xi_W &= -\rho W \pa_\rho + v\xi_{T=W}, \\
\xi_Y &=  Y^a \pa_a,\\
\end{aligned}
\label{diffeos}
\end{equation}
respectively. Next, we want to determine how the quantities defined in the previous subsection transform under the diffeomorphisms in \eqref{diffeos}. By computing the Lie derivative of the codimension-2 metric with respect to the diffeomorphisms \eqref{diffeos}, we obtain the following behaviours 
\begin{equation}
\begin{aligned}
    \var_\xi \mr{q}_{ab}&= (\tau\pa_v + \Lie_Y)\mr{q}_{ab}\\
    \var_\xi \lambda_{ab}&= (\tau\pa_v +\Lie_Y - \dot{\tau}) \lambda_{ab} +2\mr{D}_{(a}\mr{D}_{b)}\tau +4\mr{\pi}_{(b}\mr{D}_{a)}\tau\\
    \var_\xi d_{ab}
     &= (\tau\pa_v + \Lie_Y -2\dot{\tau} )d_{ab} + 2 \Bigl(\mr{D}^c \mr{K}^{(n)}_{ab}-\mr{D}_{(a}\mr{K}^{(n)c}_{b)}\Bigl)\pa_c\tau
     +2\mr{K}^{(n)c}_{(b}\mr{D}_{a)}\pa_c\tau\\
     &\quad +2\Bigl(\mr{D}_c\mr{K}^{(n)c}_{(a}  +2\mpi_c\mr{K}^{(n)c}_{(a} -\mr{D}_{(a}\mtn -\mtn\mpi_{(a} \Bigl)\pa_{b)}\tau +2\mr{K}^{(n)}_{ab}\mpi^c\pa_c\tau.
\end{aligned}
\label{sol_space}
\end{equation}
One can easily see that $\lambda_{ab}$ and $d_{ab}$ transform non-covariantly under diffeomorphisms. Performing a time derivative of the first relation in \eqref{sol_space}, one obtains the behaviour of the longitudinal extrinsic curvature under symmetry transformations, while the second relation yields the transformation rule of the transversal extrinsic curvature. Practically, one obtains
\begin{equation}
\begin{aligned}
\var_{(\tau, Y)}\mr{\sigma}^{(n)}_{ab} &=(\tau\pa_v +\Lie_Y -\dot{\tau}) \mr{\sigma}^{(n)}_{ab} + \mr{D}_{\langle a}\mr{D}_{b\rangle}\tau + 2\mr{\pi}_{\langle b}\mr{D}_{a\rangle}\tau,\\
\var_{(\tau, Y)}\mr{\sigma}^{(\ell)}_{ab} &=(\tau\pa_v +\Lie_Y +\dot{\tau}) \mr{\sigma}^{(\ell)}_{ab} ,\\
\var_{(\tau, Y)}\mr{\theta}^{(n)} &=(\tau\pa_v +\Lie_Y -\dot{\tau}) \mr{\theta}^{(n)} + \mr{D}^2\tau + 2\mr{\pi}^a\mr{D}_{a}\tau,\\
\var_{(\tau, Y)}\mr{\theta}^{(\ell)} &=(\tau\pa_v +\Lie_Y +\dot{\tau}) \mr{\theta}^{(\ell)}.
\end{aligned}
\label{tr_rule1}
\end{equation}
The behaviour of the non-affinity parameter and the Hájiček field can be easily computed by evaluating the Lie derivative of the metric component $g_{vv}$ and $g_{va}$, and read
\begin{equation}
\begin{aligned}
\var_{(\tau, Y)} \mr{\kappa}_{(\ell)} &= (\tau\pa_v +\Lie_Y+\dot{\tau} )\mr{\kappa}_{(\ell)},\\
\var_{(\tau, Y)} \mr{\pi}_a &=(\tau \pa_v +\Lie_Y  )\mr{\pi}_a + \mr{\kappa}\pa_a\tau +\pa_a \dot{\tau} - \mr{K}^{(\ell)b}_a\pa_b\tau.
\end{aligned}
\label{tr_rule2}
\end{equation}
Before delving into the construction of the covariant observables, let us spend a few words on the boost symmetry in following subsection.

\subsection{Boost symmetry}
At the beginning, we introduced a rescaling symmetry \eqref{boost_res}, under which the metric is invariant. This symmetry has been studied in \cite{Ciambelli:2023mir} and yields a non-vanishing charge on the corner. It acts infinitesimally as follows
\begin{equation}
    \var_{\lambda_L} \ell^\mu = \lambda_L \ell^\mu, \qquad \var_{\lambda_L} n^\mu = -\lambda_L n^\mu, \qquad \var_{\lambda_L} \mr{q}_{ab}=0.
\end{equation}
In particular, requiring that $\var_{(\xi, \lambda_L)}\ell=0$, one obtains that $\lambda_L(x^c)=-W(x^c)$. Previously, we showed how this symmetry acts on the extrinsic curvatures, on the Hájiček field and on the non-affinity parameter. In general, given an arbitrary quantity $\cO$, we say that $\cO$ has boost weight $w$ if transforms as follows
\begin{equation}
    \cO_{(w)}\to e^{w\lambda_L}\cO_{(w)}
\label{rule_boost}
\end{equation}
under the boost transformation in \eqref{boost_res}. Therefore, since the behaviour of the extrinsic curvatures under boost transformations obey the rule in \eqref{rule_boost}, we have the following boost weights
\begin{equation}
\begin{aligned}
w(\sigma^{(n)}_{ab})=-1, \qquad w(\sigma^{(\ell)}_{ab})=1,\qquad w(\theta^{(n)})=-1 \qquad w(\thell)=1.
\end{aligned}
\end{equation}
Conversely, $\pi_a$ and $\kappa_{(\ell)}$ do not follow the transformation rule \eqref{rule_boost} since they exhibit extra terms under the boost transformation and, therefore, are not boost-weighted quantities.

\section{Covariant observables}\label{cov_obs}
In the previous section, we briefly reviewed some basic and well-known symmetry properties of arbitrary null hyper-surfaces located at a finite distance. In particular, we observed that some quantities transform non-homogeneously under the action of the near-horizon symmetry transformations, exhibiting extra terms called the anomaly terms; these are identified through the anomaly operator \cite{Chandrasekaran:2020wwn, Freidel:2021cjp}
\begin{equation}
    \Delta_\xi\cO = (\var_\xi - \Lie_\xi)\cO,
\end{equation}
where $\cO$ is an arbitrary functional of the metric. Specifically, one can distinguish the following general structure in the transformation rules listed previously,
\begin{equation}
\var_{(\tau,Y)} \cO = [\tau\pa_v + \Lie_Y + w\dot{\tau}] \cO + L^a_\cO \pa_a \tau + \hat{L}^a_\cO \pa_a \dot{\tau} + Q^{ab}_\cO D_a \pa_b \tau + \hat{Q}^{ab}_\cO D_a \pa_b \dot{\tau},
\end{equation}
where $w$ is the boost weight. The first two inhomogeneous terms are called linear anomalies, while the last two terms are called quadratic anomalies. In the spirit of \cite{Freidel:2021qpz}, in this section we construct some quantities that transform as pseudo-tensors of weight $w$ under the near-horizon symmetry group. In other words, we define combinations of the quantities introduced in the previous section such that their quadratic anomaly and linear Weyl anomaly (i.e., $\hat{L}^a_\cO$) vanish.\\
In particular, assuming that the same pattern found at null infinity holds at a finite distance, these covariant functionals should be related to the near-horizon Weyl scalar, computed in \cite{Ruzziconi:2025fuy}. Unlike \cite{Freidel:2021qpz}, we start our treatment by analysing the transformation rule of the hypothetical boost-weighted $-2$ covariant functional $\cT_{ab}$. Once the symmetry transformation of boost-weighted $-2$ quantity is determined, the boost-weighted $-1$ covariant functional should appear in the linear anomaly of $\cT_{ab}$, and so on for the other boost-weighted $w$ functionals.

\subsection{$w=-2$ covariant functional}
Let us start the construction of the covariant functionals by searching for a quantity of boost weight $w=-2$. The sub-sub-leading contribution of the boundary metric transforms with boost weight $-2$ under \eqref{boost_res}, i.e. $d_{ab}\to e^{-2\lambda_L} d_{ab}$, but it exhibits anomalous terms under the action of the near-horizon symmetry transformations, as shown in \eqref{sol_space}. Specifically, the anomaly of $d_{ab}$ reads as follows
\begin{equation}
\begin{aligned}
\Delta_\xi d_{ab} 
&= 2 \Bigl(\mr{D}^c \mr{K}^{(n)}_{ab}-\mr{D}_{(a}\mr{K}^{(n)c}_{b)}\Bigl)\pa_c\tau +2\mr{K}^{(n)c}_{(b}\mr{D}_{a)}\pa_c\tau\\
&\quad +2\Bigl(\mr{D}_c\mr{K}^{(n)c}_{(a}  +2\mpi_c\mr{K}^{(n)c}_{(a} -\mr{D}_{(a}\mtn -\mtn\mpi_{(a} \Bigl)\pa_{b)}\tau +2\mr{K}^{(n)}_{ab}\mpi^c\pa_c\tau.
\end{aligned}
\label{anom_dab}
\end{equation}
In particular, we are looking for a quantity that transforms homogeneously under the action of the homogeneous near-horizon group. Therefore, it is convenient to distinguish in \eqref{anom_dab} a quadratic contribution, that is
\begin{equation}
\begin{aligned}
\Delta^{(2)}_\xi {d}_{\langle ab\rangle} &= 2\mr{\sigma}^{(n)}_{c\langle b} \mr{D}_{a\rangle}\pa^c\tau + \mtn \mr{D}_{\langle  a}\pa_{b\rangle}\tau\\
&= \mr{\sigma}^{(n)}_{ab} \mr{D}_{c}\pa^c\tau + \mtn \mr{D}_{\langle  a}\pa_{b\rangle}\tau,
\end{aligned}
\end{equation}
where we used the following relation \footnote{Proved in \cite{Freidel:2021qpz}.}
\begin{equation}
\mr{\sigma}_{c\langle a}^{(n)}\mr{D}_{b\rangle}\pa^c \tau = \f{1}{2}\sigma^{(n)}_{ab}\mr{D}_c\pa^c\tau,
\end{equation}
and a linear contribution
\begin{equation}
\begin{aligned}
\Delta^{(1)}_\xi d_{\langle ab\rangle} 
&= 2 \Bigl(\mr{D}_c \mr{\sigma}^{(n)}_{ab} -\mr{D}_{\langle a} \mr{K}^{(n)}_{b\rangle c} \Bigl)\pa^c\tau+2\Bigl(\mr{D}_c\mr{K}^{(n)c}_{\langle a}  + 2\mpi_c\mr{K}^{(n)c}_{\langle a} -\mr{D}_{\langle a}\mtn\\
&\qquad -\mtn\mpi_{\langle a} \Bigl)\pa_{b\rangle}\tau + 2\mr{\sigma}^{(n)}_{ab}\mpi^c\pa_c\tau.
\end{aligned}
\end{equation}
Now, using the transformation rules listed in subsection \ref{sec1.2}, it is straightforward to see that the quadratic anomaly of the trace-free part of $d_{ab}$ has the same structure as the quadratic anomaly of the following combination
\begin{equation}
\begin{aligned}
    \Delta_\xi(\mtn \mr{\sigma}_{ab}^{(n)}) &= \mtn\Delta_\xi( \mr{\sigma}_{ab}^{(n)}) + \mr{\sigma}_{ab}^{(n)}\Delta_\xi( \mtn)\\
    &=\mtn\mr{D}_{\langle a}\pa_{b\rangle} \tau +2\mtn\mpi_{\langle b} \pa_{a\rangle} \tau +\mr{\sigma}_{ab}^{(n)}\mr{D}^c\pa_c\tau + 2\mr{\sigma}_{ab}^{(n)}\mpi^c\pa_c\tau.
\end{aligned}
\end{equation}
Therefore, the quantity
\begin{equation}
    \cT_{ab} = d_{\langle ab\rangle} - \mtn \mr{\sigma}^{(n)}_{ab}
\end{equation}
possesses no quadratic anomaly, i.e. $\Delta^{(2)}_\xi \cT_{ab}=0$. In particular, it is immediate to see that the tensor $\cT_{ab}$ transforms according to the rule in \eqref{rule_boost} with boost weight $w=-2$. Putting all the previous contributions together and using the fact that \footnote{\label{note}This can be shown by introducing a 4-tensor $T_{abcd}=\mr{D}_a\mr{\sigma}^{(n)}_{bc}\pa_d\tau$, and noticing that 
\begin{equation*}
    T_{abc}^{\quad c} - T_{cab}^{\quad c} + T^{\  c}_{c\ ab} = q_{ab}T^{\ c\ d}_{c\ d},
\end{equation*}
with $T_{[abcd]}=0,\ T_{[abc]d}=0,\ T_{a[bcd]}=0, \ T_{a[bc]d}=0, \ T_{a\ cb}^{\ c}=0$.
}
\begin{equation}
\Bigl(\mr{D}_c\mr{\sigma}^{(n)}_{ab} - \mr{D}_{\langle a}\mr{\sigma}^{(n)}_{b\rangle c}\Bigl)\pa^c\tau = \mr{D}_{c}\mr{\sigma}^{(n)c}_{\langle a} \pa_{b\rangle}\tau,
\end{equation}
we obtain the transformation rule of the boost-weighted $-2$ tensor $\cT_{ab}$ under the action of the near-horizon symmetry transformations,
\begin{equation}
\begin{aligned}
\var_{(\tau,Y)}\cT_{ab} &= (\tau\pa_v +\Lie_Y -2\dot{\tau})\cT_{ab} + 4\Bigl( (\mr{D}_{c}+\mpi_c)\mr{\sigma}^{(n)c}_{\langle a} -\f{1}{2} (\mr{D}_{\langle a}+\mpi_{\langle a})\mtn \Bigl)\pa_{b \rangle}\tau\\
&= (\tau\pa_v +\Lie_Y -2\dot{\tau})\cT_{ab} + 4\cP_{\langle a} \pa_{b \rangle}\tau,
\end{aligned}
\label{var_T}
\end{equation}
where we introduced the following quantity
\begin{equation}
\cP_a = (\mr{D}_{c}+\mpi_c) \mr{\sigma}^{(n)c}_{ a} -\f{1}{2} (\mr{D}_{a} +\mpi_{a})\mtn,
\end{equation}
which denotes the linear anomaly in \eqref{var_T}. Assuming that the same pattern found at future null infinity holds at a finite distance, the equation \eqref{var_T}
suggests that $\cP_a$ must be the boost-weighted $w(\cP_a) =-1$ covariant functional. Let us analyse its properties in the next subsection.

\subsection{$w=-1$ covariant functional}
Let us now analyse the quantity that appears as a linear anomaly term in the transformation rule of $\cT_{ab}$. If one assumes that the pattern found at null infinity also holds at a finite distance, $\cP_a$ should have boost weight $w=-1$. However, we notice that $\cP_a$ contains the Hájiček field, which transforms anomalously under boost transformations. Then, before analysing the behaviour of $\cP_a$ under near-horizon diffeomorphisms, let us determine its transformation under \eqref{boost_res}. We have
\begin{equation}
\begin{aligned}
\cP_a &\to \mr{D}_b(e^{-\lambda_L}\mr{\sigma}^{(n)b}_a) + e^{-\lambda_L} \mpi_b\mr{\sigma}^{(n)b}_a + e^{-\lambda_L} \mr{\sigma}^{(n)b}_a \pa_b\lambda_L - \f{1}{2}\mr{D}_a(e^{-\lambda_L}\mtn)\\
&\qquad- \f{1}{2}e^{-\lambda_L}\mpi_a\mtn - \f{1}{2}e^{-\lambda_L}\mtn \pa_a\lambda_L\\
&=e^{-\lambda_L}\cP_a ,
\end{aligned}
\end{equation}
and therefore $\cP_a$ transforms according to the rule in \eqref{rule_boost} with boost weight $w=-1$. Now, in order to analyse its behaviour under the near-horizon symmetry transformations, it might be convenient to rewrite $\cP_a$ as follows
\begin{equation}
\cP_a = (\mr{D}_{b}+\mpi_b)\mr{K}^{(n)b}_{a} - (\mr{D}_{a}+\mpi_{a})\mtn.\label{cpa}
\end{equation}
Using the results in \eqref{tr_rule1}-\eqref{tr_rule2} and the relations \eqref{Gamma_anom} in appendix \ref{formulae}, we obtain that the first term in \eqref{cpa} possesses the following anomaly structure
\begin{equation}
\begin{aligned}
\Delta_\xi(\mr{D}_b \mr{K}^{(n)b}_{a})
&= \pa_v\mr{K}^{(n)b}_a \pa_b\tau - \mr{K}^{(n)b}_a \pa_b\dot{\tau} + \f{1}{2}\mr{\cR} \pa_a \tau +\mr{D}_a \mr{D}^2\tau -2\mr{K}_{(n)}^{bc}\mr{K}^{(\ell)}_{c(a} \pa_{b)}\tau\\
&\quad +\mr{D}_b\mpi^{b} \pa_{a}\tau +\mr{D}_b\mpi_{a}\pa^{b}\tau+\mpi^{b}\mr{D}_a\mr{D}_{b}\tau+\mpi_{a}\mr{D}^2\tau
\\
&\quad +\mr{K}^{(n)b}_{a}\mthell\pa_b\tau +  \mr{K}^{(n)b}_{c} \mr{K}^{(\ell)}_{ab}\pa^c\tau,
\end{aligned}
\end{equation}
and
\begin{equation}
\begin{aligned}
\Delta_\xi(\mpi_b \mr{K}^{(n)b}_{a}) &= \mr{K}^{(n)b}_{a}\mr{\kappa}\pa_b\tau +\mr{K}^{(n)b}_{a}\pa_b \dot{\tau} - \mr{K}^{(n)b}_{a}\mr{K}^{(\ell)c}_b\pa_c\tau + \mpi^b\mr{D}_{(a}\mr{D}_{b)}\tau \\
&\qquad +2 \mpi^b\mpi_{(b}\mr{D}_{a)}\tau.
\end{aligned}
\end{equation}
Additionally, the last two terms in \eqref{cpa} easily read as follows
\begin{equation}
\begin{aligned}
\Delta_\xi[(\mr{D}_a +\mpi_a)\mtn] &= [(\pa_v +\mkell)\mtn ]\pa_a\tau  + \mr{D}_a\mr{D}^2\tau +2\mr{D}_a\mpi^b\pa_b\tau\\
&+2\mpi^b\mr{D}_a\pa_b\tau +\mpi_a\mr{D}^2\tau +2\mpi_a\mpi^b\pa_b\tau - \mtn\mr{K}^{(\ell)b}_a\pa_b\tau.
\end{aligned}
\end{equation}
Putting all of the above contributions together, we have that the quadratic anomaly of the boost $w=-1$ metric functional \eqref{cpa} vanishes, i.e. $\Delta^{(2)}_\xi \cP_a=0$, and also the linear Weyl anomalies (those $\propto \pa_a\dot{\tau}$) cancel each other out. Then, we are left with the following linear anomaly
\begin{equation}
\begin{aligned}
\Delta_\xi \cP_a &= (\pa_v + \mr{\kappa})\mr{K}^{(n)b}_a\pa_b\tau  + \f{1}{2}\mr{\cR} \pa_a \tau -2\mr{K}_{(n)}^{bc}\mr{K}^{(\ell)}_{c(a} \pa_{b)}\tau +\mr{D}_b\mpi^{b} \pa_{a}\tau +\mr{D}_b\mpi_{a}\pa^{b}\tau\\
&\qquad +\mr{K}^{(n)b}_{a}\mthell\pa_b\tau +  \mr{K}^{(n)b}_{c} \mr{K}^{(\ell)}_{ab}\pa^c\tau
 - \mr{K}^{(n)b}_{a}\mr{K}^{(\ell)c}_b\pa_c\tau   
+ \mpi^b\mpi_{b}\pa_{a}\tau\\
&\qquad-(\pa_v +\mkell)\mtn \pa_a\tau   -2\mr{D}_a\mpi^b\pa_b\tau
-\mpi_a\mpi^b\pa_b\tau+ \mtn\mr{K}^{(\ell)b}_a\pa_b\tau.
\end{aligned}
\end{equation}
Using the fact that $\mr{\cR}_{ab}[\mr{q}]=\f{1}{2}\mr{q}_{ab}\mr{\cR}[\mr{q}]$ and the following relation
\begin{equation}
2\sigma_{c(a}^{(n)}\sigma^{(\ell)c}_{b)} =\mr{q}_{ab} \sigma_{cd}^{(n)}\sigma_{(\ell)}^{cd},
\end{equation}
after some algebraic manipulations the linear anomaly of $\cP_a$ can be put in the following form
\begin{equation}
\begin{aligned}
    \Delta_\xi\cP_a &= \f{1}{2}\mr{\mathbb{E}}_a^{\ b}\pa_b\tau - \f{1}{2}\mr{\mathbb{E}}_b^{\ b}\pa_a\tau + \f{3}{2}\Bigl(\mr{\cR}_{ab} + \f{1}{2}\mtn\mthell \mr{q}_{ab}-  2\mr{\sigma}^{(n)}_{c(a} \mr{\sigma}_{b)}^{(\ell)c} \Bigl)\pa^b\tau\\
    &\quad -3 \mr{\sigma}^{(n)}_{c[a} \mr{\sigma}_{b]}^{(\ell)c}\pa^b\tau  - 3\mr{D}_{[a}\mpi_{b]}\pa^b\tau\\
    &\heq 3\cA_{ab}\pa^b\tau,
\end{aligned}
\end{equation}
where
\begin{equation}
\cA_{ab} = \f{1}{2}\Bigl(\mr{\cR}_{ab} + \f{1}{2}\mtn\mthell \mr{q}_{ab}-  2\mr{\sigma}^{(n)}_{c(a} \mr{\sigma}_{b)}^{(\ell)c} \Bigl) - \mr{\sigma}^{(n)}_{c[a} \mr{\sigma}_{b]}^{(\ell)c}  - \mr{D}_{[a}\mpi_{b]}.
\end{equation}
Thus, the behaviour under the near-horizon symmetry transformations of the boost-weighted functional $\cP_a$ is
\begin{equation}
\begin{aligned}
\var_{(\tau, Y)}\cP_a &\heq (\tau\pa_v + \Lie_Y - \dot{\tau}) \cP_a + 3\cA_{ab}\pa^b\tau.\\
\end{aligned}
\end{equation}
From the transformation rule introduced in section \ref{review1}, it is very straightforward to notice that the quantity $\cA_{ab}$ has boost weight equal to zero.

\subsection{$w=0$ covariant functional}\label{area_dual}
Before delving into the analysis of the behaviour of the boost-weight-0 covariant functional under symmetry transformations, let us introduce the duality symmetry. Given two arbitrary quantities $Y_a$ and $X_{ab}$ on $\cS$, their duals are defined as follows
\begin{equation}
\wt{Y}_a =\varepsilon_a^{\ b} Y_b, \qquad \wt{X}_{ab} = \varepsilon_a^{\ c} X_{cb},
\label{duality}
\end{equation}
where $\varepsilon_{ab}$ is the volume form on $\cS$, i.e. $\epsilon_{ab} \sqrt{\mr{q}} =\varepsilon_{ab}$, also defined in appendix \ref{NP_an} via the complex dyad. 
In particular, from the following relation $\varepsilon_{a}^{\ c} \varepsilon_b^{\ a} = -\var_b^{\ c}$, one can infer that the dual of a dual 1-form is equal to minus the 1-form itself, namely ${\wt{\wt{Y}}}_a=\varepsilon_a^{\ b} \wt{Y}_b=-Y_a$. The tensorial density $\varepsilon_{ab}$ transforms as follows under symmetry transformations
\begin{equation}
\var_{(\tau,Y)} \varepsilon_{ab}= (\tau\pa_v + \mr{D}_cY^c)\sqrt{\mr{q}}\ \epsilon_{ab},
\end{equation}
and the mixed tensor is defined as $\varepsilon_a^{\ b} :=\varepsilon_{ac}q^{cb}$. Having introduced the duality transformation, let us now analyse the behaviour of the boost-0 quantity
\begin{equation}
\cA_{ab} = \f{1}{2}\Bigl(\mr{\cR}_{ab} + \f{1}{2}\mtn\mthell \mr{q}_{ab}-  2\mr{\sigma}^{(n)}_{c(a} \mr{\sigma}_{b)}^{(\ell)c} \Bigl) - \mr{\sigma}^{(n)}_{c[a} \mr{\sigma}_{b]}^{(\ell)c}  - \mr{D}_{[a}\mpi_{b]}
\end{equation}
under symmetry transformations. In order to simplify the calculations, it is useful to evaluate the symmetric and the anti-symmetric parts of $\cA_{ab}$ separately, i.e.,
\begin{equation}
\cA_{ab} = \cA_{(ab)} +  \cA_{[ab]}.
\end{equation}
The symmetric part of $\cA_{ab}$ reads as follows
\begin{equation}
\cA_{(ab)}= \f{1}{4} \Bigl(\mthell\mtn +\mr{\cR}-2\mr{\sigma}^{(n)}_{cd} \mr{\sigma}^{cd}_{(\ell)}\Bigl) \mr{q}_{ab}=\f{1}{2}\cA \mr{q}_{ab},
\label{A(ab)}
\end{equation}
and using the transformations rule provided in section \ref{review1} and in appendix \ref{formulae}, we have
\begin{equation}
\begin{aligned}
\Delta_\xi(\thell\mtn-2\mr{\sigma}^{(n)}_{ab} \mr{\sigma}_{(\ell)}^{ab}) &= \mthell\mr{D}^c\pa_c \tau+2\mthell \mr{\pi}^c\pa_c\tau-2\mr{\sigma}_{(\ell)}^{ab}\mr{D}_{\langle a}\pa_{b\rangle}\tau \\
&\quad - 4 \mr{\sigma}_{(\ell)}^{ab} \mr{\pi}_{\langle b}\pa_{a\rangle}\tau,
\end{aligned}
\end{equation}
and
\begin{equation}
\begin{aligned}
\Delta_\xi\mr{\cR} &= 4\mr{D}_{a} \mr{\sigma}_{(\ell)}^{ab}\pa_b\tau-2\mr{D}^c \mthell \pa_c\tau + 2\mr{\sigma}_{(\ell)}^{ab} \mr{D}_{\langle a} \pa_{b\rangle} \tau - \mthell\mr{D}^c\pa_c \tau.
\end{aligned}
\end{equation}
The quadratic anomaly terms cancel each other out, and we are left with the following transformation rule
\begin{equation}
\begin{aligned}
\var_\xi \cA &=(\tau\pa_v + \Lie_Y)\cA + 2\Bigl((\mr{D}_{a} -\mpi_{ a}) \mr{\sigma}_{(\ell)}^{ab} - \f{1}{2}(\mr{D}^{b} -\mpi^{b}) \mthell \Bigl)\pa_b\tau  \\
&=(\tau\pa_v + \Lie_Y)\cA  + 2\cJ^a \pa_a \tau,
\end{aligned}
\end{equation}
where we denoted by $\cJ^a$ the linear anomaly contribution enclosed in the round bracket. Now, let us focus on the anti-symmetric part of $\cA_{ab}$, which can be written as follows
\begin{equation}
\cA_{[ab]} = -\f{1}{2}\Bigl(\mr{\sigma}^{(n)d}_{c} \mr{\wt{\sigma}}^{(\ell)c}_{d}  + \mr{\wt{D}}_{c}\mpi^{c}\Bigl)\varepsilon_{ab} = \f{1}{2}\wt{\cA}\varepsilon_{ab},
\label{A[ab]}
\end{equation}
where we used the duality transformation in \eqref{duality} and we introduced the scalar $\wt{\cA}$ as the quantity enclosed in the round bracket. Under the action of the near-horizon symmetry group, we have that 
\begin{equation}
\begin{aligned}
\Delta_\xi(\varepsilon^{ab}\mr{D}_a\mpi_b) &= \varepsilon^{ab}( \pa_v\mpi_b -  \pa_b\mr{\kappa}_{(\ell)}) \pa_a\tau -\varepsilon^{ab}\mr{D}_a\mr{K}^{(\ell)c}_b\pa_c\tau - \varepsilon^{ab} \mr{\sigma}^{(\ell)c}_b \mr{D}_a\pa_c\tau \\
&\heq  \varepsilon^{ab}( \pa_a\mthell -\mthell\mpi_a -2 \mr{D}_c \mr{\sigma}^{(n)c}_a ) \pa_b\tau - \varepsilon^{ab} \mr{\sigma}^{(\ell)c}_b \mr{D}_a\pa_c\tau ,\\
\end{aligned}    
\end{equation}
where we imposed the Damour equation \eqref{damour} and used the following relation
\begin{equation}
\mr{D}_{[a}\mr{\sigma}^{(n)c}_{b]}\pa_c \tau = \mr{D}_{c} \mr{\sigma}^{(n)c}_{[a}\pa_{b]} \tau,
\end{equation}
that comes from the identity $\mr{D}_{[a}\mr{\sigma}^{(n)c}_{b}\pa_{c]} \tau=0$. The other term reads
\begin{equation}
\begin{aligned}
\Delta_\xi (\mr{\sigma}^{(n)a}_{b} \mr{\wt{\sigma}}^{(\ell)b}_{a} ) &= \mr{\wt{\sigma}}^{ab}_{(\ell)} \mr{D}_{\langle a} \pa_{b\rangle}\tau + 2\mr{\wt{\sigma}}^{ab}_{(\ell)}\mpi_{\langle a} \pa_{b\rangle}\tau.
\end{aligned}
\end{equation}
Putting the previous contributions together into the expression of $\wt{\cA}$, we finally obtain
\begin{equation}
\begin{aligned}
    \var_\xi \wt{\cA} &= (\tau\pa_v + \Lie_Y -\dot{\tau}) \wt{\cA}  + 2\varepsilon^{ab}\Bigl((\mr{D}_c-\mpi_c)\mr{\sigma}^{(\ell)c}_b - \f{1}{2}(\mr{D}_b-\mpi_b)\mthell \Bigl) \pa_a \tau\\
    &= (\tau\pa_v + \Lie_Y -\dot{\tau}) \wt{\cA}  + 2\wt{\cJ}^a \pa_a \tau,
\end{aligned}
\end{equation}
where we defined $\wt{\cJ}^a = \varepsilon^{ab}\cJ_a$. Because it shares the same structure as the boost-weight $-1$ functional, it is straightforward to see that $\cJ^a$ transforms covariantly under \eqref{boost_res} with boost weight equal to 1, i.e.
\begin{equation}
\cJ_a \to e^{\lambda_L}  \cJ_a.  
\end{equation}
In particular, since $\varepsilon^{ab}$ is invariant under \eqref{boost_res}, the dual quantity $\wt{\cJ}^a$ transforms as $\cJ_a$ under boost transformations.

\subsection{$w=1,2$ covariant functionals}
The behaviour of the zero-boost-weight covariant functional under symmetry transformations suggests the form of the boost-weight 1 functional $\cJ_a$, which reads
\begin{equation}
\begin{aligned}
\cJ_a &= (\mr{D}_b - \mpi_b) \mr{\sigma}_a^{(\ell)b}  - \f{1}{2}(\mr{D}_a -\mpi_a)\mthell\\
&= (\mr{D}_b - \mpi_b) \mr{K}_a^{(\ell)b}  - (\mr{D}_a -\mpi_a)\mthell.
\end{aligned}
\label{ja}
\end{equation}
Let us analyse how this quantity transforms under the near-horizon symmetry group. The first term in \eqref{ja} transforms according to the following rule
\begin{equation}
\begin{aligned}
\Delta_\xi(\mr{D}_b\mr{K}^{(\ell)b}_{a}) &= \pa_v\mr{K}^{(\ell)b}_{a} \pa_b\tau + \mr{K}^{(\ell)b}_{a}\pa_b\dot{\tau} +\mthell \mr{K}^{(\ell)b}_{a} \pa_b\tau -2\mr{K}^{(\ell)b}_c \mr{K}^{(\ell)c}_{(a} \pa_{b)} \tau\\
&\quad + \mr{K}^{(\ell)b}_c \mr{K}^{(\ell)}_{ab} \pa^c\tau,
\end{aligned}
\end{equation}
and
\begin{equation}
\begin{aligned}
\Delta_\xi(\mpi_b\mr{K}^{(\ell)b}_{a}) &= \mr{K}^{(\ell)b}_{a}\mr{\kappa}\pa_b\tau +\mr{K}^{(\ell)b}_{a}\pa_b \dot{\tau} -\mr{K}^{(\ell)b}_{a}\mr{K}^{(\ell)c}_b\pa_c\tau.
\end{aligned}
\end{equation}
Then, using the transformation rules listed in subsection \ref{sec1.2} and putting all the contributions together, we obtain the following linear anomaly for $\cJ_a$,
\begin{equation}
\begin{aligned}
\Delta_\xi \cJ_a &\heq (\pa_v\mr{K}_a^{(\ell)b} - \mr{\kappa}_{(\ell)}\mr{K}_a^{(\ell)b} + \mr{K}_a^{(\ell)c}\mr{K}_c^{(\ell)b} )\pa_b\tau -\mr{\mathbb{E}}_{\ell\ell}\pa_a \tau,
\end{aligned}
\end{equation}
where we used the null Raychaudhuri equation \eqref{null_ray}. Hence, one can infer that $\cJ_a$ transforms as follows under the near-horizon symmetry group
\begin{equation}
\var_\xi \cJ_a = (\tau\pa_v +\Lie_Y + \dot{\tau}) \cJ_a + \cN_a^{\ b}\pa_b\tau,
\end{equation}
where we identify the covariant functional of boost weight 2 as
\begin{equation}
\cN_{ab} = \pa_v\mr{K}_{ab}^{(\ell)} - \mr{\kappa}_{(\ell)}\mr{K}_{ab}^{(\ell)} - \mr{K}_a^{(\ell)c}\mr{K}_{cb}^{(\ell)},
\end{equation}
which is indeed anomaly-free, i.e. $\Delta_\xi \cN_{ab}=0$, and transforms as follows
\begin{equation}
\var_\xi\cN_{ab} = (\tau\pa_v + \Lie_Y +2\dot{\tau}) \cN_{ab}.
\end{equation}
In particular, the trace of $\cN_{ab}$ vanishes on shell, indeed $ \cN_a^{\ a} = \mr{\mathbb{E}}_{\ell\ell}$ and therefore 
\begin{equation}
\cN_{ab} \heq\cN_{\langle ab\rangle} = (\pa_v\mr{\sigma}_{ab}^{(\ell)} )^\TF - \mr{\kappa}_{(\ell)}\mr{\sigma}_{ab}^{(\ell)}.
\end{equation}
In spite of the presence of the non-affinity parameter $\mkell$ in the expression of $\cN_{ab}$, which does not transform homogeneously under boost transformations, the full $\cN_{ab}$ obeys the transformation rule in \eqref{rule_boost}, indeed
\begin{equation}
\begin{aligned}
\cN_{ab} &\to e^{\lambda_L} \pa_v(e^{\lambda_L}\mr{K}^{(\ell)}_{ab}) -e^{2\lambda_L}\mkell\sigell_{ab} -e^{2\lambda_L}\mr{K}^{(\ell)}_{ab} \pa_v\lambda_L -e^{2\lambda_L} \mr{K}^{(\ell)}_{ac} \mr{K}^{(\ell)c}_{b} \\
&=e^{2\lambda_L}\cN_{ab}.
\end{aligned}
\end{equation}

\section{Evolution equations}
In the previous discussion, we determined a set of boost-weighted covariant functionals $\mathrm{Q}_w= (\cN_{ab}, \cJ_a, \cA_{ab}, \cP_a, \cT_{ab})$ just by constructing a homogeneous tensor out of the trace-free part of $d_{ab}$ and showing that the following pattern
\begin{equation}
\var_{(\tau,Y)}\mathrm{Q}_w= (\tau\pa_v + \Lie_Y + w\dot{\tau})\mathrm{Q}_w - (w-2)\mathrm{Q}_{w+1}\pa\tau
\end{equation}
holds. In other words, the knowledge of the lowest boost-weighted functional ($w^h=-2$ in our case) allows the construction of the boost weighted $w=w^h+1,...,2$ functionals.\\
Now, we would like to evaluate the near-horizon evolution equations of the covariant functionals defined above and the properties of the former under symmetry transformations. Let us consider the time derivative of the covariant current, knowing that
\begin{equation}
\pa_v\cJ_a = \pa_v\Bigl[(\mr{D}_b - \mpi_b) \mr{\sigma}_a^{(\ell)b}-\f{1}{2}(\mr{D}_a -\mpi_a)\mthell\Bigl].    
\label{pav_J}
\end{equation}
Thus, let us evaluate the rhs of the equation \eqref{pav_J}. In particular, given an arbitrary tensor $T^{\ b}_{a}$, we use the following relation
\begin{equation}
\begin{aligned}
\pa_v\mr{D}_b(T^{\ b}_{a}) &=\pa_v(\pa_b T^{\ b}_{a} + \mr{\Gamma}^b_{bc} T_a^{\ c} - \mr{\Gamma}^c_{ab}T^{\ b}_c)\\
&=\mr{D}_b\pa_v T^{\ b}_{a} + T^b_{\ a} \mr{D}_b\mthell - T^b_{\ c}(\mr{D}_a \mr{K}^{(\ell)c}_{b} +\mr{D}_b\mr{K}^{(\ell)c}_{a}-\mr{D}^c\mr{K}^{(\ell)}_{ab}),
\end{aligned}
\label{pav_A}
\end{equation}
where
\begin{equation}
\begin{aligned}
\pa_v\mr{\Gamma}^c_{ab} &= (\mr{D}_a \mr{K}^{(\ell)c}_{b} +\mr{D}_b\mr{K}^{(\ell)c}_{a}-\mr{D}^c\mr{K}^{(\ell)}_{ab})\\
\end{aligned}
\end{equation}
and $\pa_v\Gamma^b_{bc} =\mr{D}_c\mthell$. Then, we have 
\begin{equation}
\begin{aligned}
\pa_v\cJ_a &= \mr{D}_b\pa_v K^{(\ell) b}_{a} + K^{(\ell)b}_{\ a} \mr{D}_b\mthell - K^{(\ell)b}_{\ c}(\mr{D}_a \mr{K}^{(\ell)c}_{b} +\mr{D}_b\mr{K}^{(\ell)c}_{a}-\mr{D}^c\mr{K}^{(\ell)}_{ab})\\
&\quad -\mpi_b\pa_vK^{(\ell) b}_{a} -K^{(\ell) b}_{a}\pa_v\mpi_b - \mr{D}_a\pa_v\mthell + \mpi_a\pa_v\mthell+ \mthell\pa_v\mpi_a.
\end{aligned}
\end{equation}
and using the null Raychaudhuri equation \eqref{null_ray} and Damour equation \eqref{damour} in the following form
\begin{equation}
\mr{\mathbb{E}}_{\ell a} =\pa_v\mpi_a + \mpi_b\mr{K}^{(\ell)b}_a - \pa_a\mkell + \cJ_a
\end{equation}
after some algebraic manipulations we obtain
\begin{equation}
\begin{aligned}
    \pa_v\cJ_a &= \Bigl(\mkell -2\mthell\Bigl)\cJ_a + \mr{K}^{(\ell)b}_{a} \cJ_b + (\mr{D}_b -\mpi_b)\cN^{\ b}_a -\mr{K}_a^{(\ell)b}\mr{\mathbb{E}}_{\ell b}\\
    &\quad -(\mr{D}_a - \mpi_a) \mr{\mathbb{E}}_{\ell\ell} + \mthell\mr{\mathbb{E}}_{\ell a}.
\end{aligned}
\end{equation}
Therefore, imposing the null Raychaudhuri and Damour equations to hold, we find the evolution equation of the boost-1 covariant functional, which reads
\begin{equation}
    \cE_\cJ := \pa_v\cJ_a -\Bigl(\mkell -2\mthell\Bigl)\cJ_a - \mr{K}^{(\ell)b}_{a} \cJ_b - (\mr{D}_b -\mpi_b)\cN^{\ b}_a.
\label{EJ}
\end{equation}
As we show in appendix \ref{ev_BI}, the evolution equation of $\cJ_a$ comes from a re-arrangement of the evolution equation of the leading order of Weyl scalar $\Psi_1$. In particular, under the action of the symmetry transformations we have
\begin{equation}
\begin{aligned}
\var_\xi(\pa_v\cJ_a)&= (\tau\pa_v +\Lie_Y + 2\dot{\tau}) \pa_v\cJ_a + \pa_v(\cN_a^{\ b}\pa_b\tau),\\
\var_\xi(\mr{D}_b\cN^{\ b}_{a}) &= (\tau\pa_v + \Lie_Y +2\dot{\tau})\cN^{\ b}_{a} + \pa_v\cN^{\ b}_{a}\pa_b\tau +2\cN^{\ b}_{a}\pa_b\dot{\tau} + \mthell\cN^{\ b}_{a}\pa_b\tau \\
&\quad -2\cN^{\ b}_{c}\mr{K}^{(\ell)c}_{(b} \mr{D}_{a)} \tau + \cN^{\ b}_{ c}\mr{K}^{(\ell)}_{ab}\mr{D}^c \tau, \\
\var_\xi(\mpi_b\cN^{\ b}_{a}) &= (\tau\pa_v + \Lie_Y +2\dot{\tau}) (\mpi_b\cN^{\ b}_{a}) + \mr{\kappa}_{(\ell)}\cN^{\ b}_{a}\pa_b\tau +\cN^{\ b}_{a}\pa_b\dot{\tau} - \cN^{\ b}_{a}\mr{K}_{\ b}^{(\ell)c} \pa_c\tau,
\end{aligned}
\end{equation}
and using the following formula
\begin{equation}
2\sigma^{(\ell)}_{c(a}\cN^{\ c}_{b)} = \mr{q}_{ab}\sigma^{(\ell)}_{cd}\cN^{cd},
\end{equation}
we obtain that $\cE_\cJ$ transforms without anomalies under the symmetry transformations, i.e.
\begin{equation}
\begin{aligned}
    \delta_{(\tau, Y)} \cE_\cJ = (\tau\pa_v +\Lie_Y + 2\dot{\tau}) \cE_\cJ.
\end{aligned}
\end{equation}
In order to identify the evolution equations of the remaining covariant functionals, one can follow the same strategy discussed in \cite{Freidel:2021qpz}. In other words, one can construct combinations of the previously introduced quantities such that the given combination transforms covariantly under near-horizon symmetries.\\

\subsection{Evolution equation for $w=0$}
Assuming that the same pattern found at null infinity holds at a finite distance (where the covariant derivative is now replaced by a Weyl connection \cite{Ciambelli:2018wre, Ciambelli:2018ojf, Ciambelli:2019bzz}), we obtain that the following combination
\begin{equation}
\cE^{\cA} :=\pa_v\cA +\f{3}{2}\mthell\cA - (\mr{D}_{a} + \mpi_{a})\cJ^{a} + \mr{\sigma}^{(n)}_{ab} \cN^{ab}
\label{cA}
\end{equation}
transforms (on-shell) covariantly under the action of symmetry transformations.  To demonstrate that the combination \eqref{cA} transforms covariantly under near-horizon symmetry transformations, let us notice that 
\begin{equation}
\begin{aligned}
\Delta_\xi (\mr{D}_a\cJ^a) &= \pa_v(\cJ^a \pa_a \tau) + \mr{D}_b(\cN^{ab}\pa_a\tau)- \cJ^a\mthell\pa_a\tau,\\
\var_\xi (\mpi_a\cJ^a) &= \mr{\kappa}_{(\ell)} \cJ^a\pa_a\tau + \cJ^a\pa_a\dot{\tau} - \cJ^a\mr{K}^{(\ell)b}_a\pa_b\tau +\mpi_a\cN^{ab} \pa_b\tau,\\
\var_\xi (\mr{\sigma}^{(n)}_{ab}\cN^{ab}) &=\cN^{ab} \mr{D}_{\langle a}\pa_{b\rangle}\tau +2\cN^{ab}\mpi_{\langle a}\pa_{b\rangle}\tau.
\end{aligned}
\end{equation}
Thus, combining the previous contributions, we obtain that the behaviour of the evolution equation of $\cA$ under symmetry transformations is
\begin{equation}
\begin{aligned}
    \var_{(\tau, Y)}\cE^\cA &=(\tau\pa_v +\Lie_Y +\dot{\tau})\cE^\cA + \Bigl[(\pa_v +2\mthell-\mkell)\cJ^a +\mr{K}^{(\ell)a}_b\cJ^b - (\mr{D}_b - \mpi_b)\cN^{ab}\Bigl]\pa_a\tau\\
    &= (\tau\pa_v +\Lie_Y +\dot{\tau})\cE^\cA + \cE^a_\cJ\pa_a\tau.
\end{aligned}
\end{equation}
Therefore, the linear anomaly $\Delta_\xi \cE^{\cA} = \cE^a_\cJ \pa_a\tau$ vanishes when $\cE_\cJ^a$ is imposed. Similarly, exploiting the duality symmetry, it is immediate to compute the evolution equation of the antisymmetric part of $\cA_{ab}$, which reads
\begin{equation}
   \cE^{\wt{\cA}} :=\pa_v\wt{\cA} +\f{3}{2}\mthell \wt{\cA} - (\mr{D}_a + \mpi_a)\wt{\cJ}^a + \mr{\sigma}^{(n)}_{ab}\wt{\cN}^{ab} ,
\end{equation}
and transforms as follows under the action of the near-horizon symmetries
\begin{equation}
    \var_{(\tau, Y)}\cE^{\wt{\cA}} = (\tau\pa_v +\Lie_Y +\dot{\tau})\cE^{\wt{\cA}} + \cE^a_{\wt{\cJ}}\pa_a\tau.
\end{equation}
Similarly, the linear anomaly $\Delta_\xi \cE^{\wt{\cA}} = \cE^a_{\wt{\cJ}} \pa_a\tau$ vanishes when $\cE_{\wt{\cJ}}^a$ is imposed.

\subsection{Evolution equation for $w=-1$}
Using the same strategy, let us now focus our attention on the evolution equation of the boost-$-1$ functional $\cP_a$. Then, we obtain the following combination
\begin{equation}
\begin{aligned}
\cE^\cP_{a} :=\pa_v\cP_a + \mr{\mu}_{(\ell)}\cP_a -\mr{\sigma}^{(\ell)b}_{a}\cP_b  -\f{1}{2}(\pa_a +3\mpi_a)\cA - \f{1}{2}(\wt{\pa}_a +3\mr{\wt{\pi}}_a)\wt{\cA} +2\mr{\sigma}^{(n)}_{ab}\cJ^b ,
\end{aligned}
\label{EP}
\end{equation}
where $\mr{\mu}_{\ell} = \mkell + \mthell/2$ is the surface tension of $\cH$. Following the transformation rules listed below
\begin{equation}
\begin{aligned}
\Delta_\xi[(\pa_v+\mr{\mu}_{(\ell)})\cP_a ]&= 3(\pa_v+\mr{\mu}_{(\ell)})(\cA_{ab} \pa^b\tau),\\
\Delta_\xi(\mr{\sigma}^{(\ell)b}_a\cP_b) &= 3\cA_{bc}\mr{\sigma}^{(\ell)b}_a \pa^c\tau,\\
\Delta_\xi(\pa_a\cA) &= \pa_v\cA\pa_a\tau + 2\mr{D}_a(\cJ^b \pa_b \tau) ,\\
\Delta_\xi(\mpi_a\cA) &= \mr{\kappa}\cA\pa_a\tau + \cA\pa_a\dot{\tau} - \cA\mr{K}^{(\ell)b}_a \pa_b\tau+ 2\mpi_a\cJ^b \pa_b \tau, \\
\Delta_\xi(\mr{\sigma}^{(n)}_{ab}\cJ^b) &= \mr{\sigma}^{(n)}_{ab}\cN^{bc}\pa_c\tau + \cJ^b\mr{D}_{\langle a}\pa_{b\rangle}\tau + 2\cJ^b\mr{\pi}_{\langle a}\pa_{b\rangle}\tau,
\end{aligned}
\end{equation}
and putting them together into the \eqref{EP}, the quadratic anomaly of $\cE^\cP_a$ can be written as follows 
\begin{equation}
\begin{aligned}
\Delta^{(2)}_\xi\cE^\cP_a &=2\cJ^b \mr{D}_{\langle a}\pa_{b\rangle}\tau -\cJ^b\mr{D}_a\pa_b\tau - \wt{\cJ}^b\mr{\wt{D}}_a\pa_b\tau,
\end{aligned}
\end{equation}
while the linear anomaly reads
\begin{equation}
\begin{aligned}
\Delta^{(1)}_\xi\cE^\cP_a &= \Bigl(\pa_v (\cA+\wt{\cA})  +\f{3}{2}\mthell (\cA+\wt{\cA})\Bigl)\pa_a\tau - \mr{D}_a\cJ^b\pa_b\tau  - \mr{\wt{D}}_a\wt{\cJ}^b \pa_b\tau + 4\cJ^b\mpi_{\langle a}\pa_{b\rangle}\tau \\
&\qquad-3\mpi_a\cJ^b\pa_b\tau - 3\mr{\wt{\pi}}_a\wt{\cJ}^b\pa_b\tau +2\mr{\sigma}^{(n)}_{ab} \cN^{bc} \pa_c\tau.
\end{aligned}
\end{equation}
Then, by exploiting the properties of the duality symmetry we have that $\wt{\cJ}^b \mr{\wt{D}}_a\pa_b\tau = -\cJ^b \mr{\wt{D}}_{a} \wt{\pa}_{b}\tau$ and using the following relation $\varepsilon_{ab} \varepsilon^{cd} = 2\var_a^{\ [c}\var_b^{\ d]}$, we obtain
\begin{equation}
\begin{aligned}
\wt{\cJ}^b \mr{\wt{D}}_a\pa_b\tau &= -\cJ^b \mr{\wt{D}}_{\langle a} \wt{\pa}_{b\rangle}\tau +\f{1}{2}\wt{\cJ}^b\varepsilon_{ab}\mr{D}^2\tau = \cJ^b \mr{D}_{\langle a} \pa_{b\rangle}\tau -\f{1}{2}\cJ_a \mr{D}^2\tau,
\end{aligned}
\end{equation}
which implies that $\Delta^{(2)}_\xi\cE^\cP_a=0$. Moreover, from the same computation the following identities also hold
\begin{equation}
\begin{aligned}
(\mr{D}_a\cJ_b +\mr{\wt{D}}_a\wt{\cJ}_b)\pa^b\tau &= (\mr{D}_c\cJ^c)\pa_a\tau + (\mr{D}_c \wt{\cJ}^c) \wt{\pa}_a\tau,\\
(\mpi_a\cJ_b +\mr{\wt{\pi}}_a \wt{\cJ}_b) \pa^b\tau&=(\mr{\pi}_c \cJ^c)\pa_a\tau + (\mr{{\pi}}_c \wt{\cJ}^c) \wt{\pa}_a\tau,\\
\end{aligned}
\end{equation}
and
\begin{equation}
\begin{aligned}
\cJ^b\mpi_{\langle a}\pa_{b\rangle}\tau 
&= \f{1}{2}\cJ^b\mpi_{b}\pa_{a}\tau + \f{1}{2}\mpi_b \wt{\cJ}^{b} \wt{\pa}_a\tau.
\end{aligned}
\end{equation}
Finally, by employing this last identity
\begin{equation}
\begin{aligned}
\mr{\sigma}^{(n)}_{ac}\cN_b^{\ c} &= \mr{\sigma}^{(n)}_{c(a}\cN_{b)}^{\ c} +\mr{\sigma}^{(n)}_{c[a}\cN_{b]}^{\ c}\\
&= \f{1}{2} \mr{q}_{ab} \mr{\sigma}^{(n)}_{cd} \cN^{cd} + \f{1}{2} \varepsilon_{ab}\mr{\sigma}^{(n)}_{cd}\wt{\cN}^{cd},
\end{aligned}
\end{equation}
we obtain that the behaviour of the equation \eqref{EP} under the action of the near-horizon symmetry is the following
\begin{equation}
\begin{aligned}
\var_{(\tau, Y)}\cE^\cP_{a}&= (\tau\pa_v +\Lie_Y )\cE_{a}^\cP+\cE^\cA\pa_a\tau +\cE^{\wt{\cA}}\wt{\pa}_a\tau \\
&= (\tau\pa_v +\Lie_Y)\cE_{a}^\cP + 2 \cE^\cA_{ab}\pa^b\tau,
\end{aligned}
\end{equation}
and its linear anomaly vanishes when $\cE^\cA$ and $\cE^{\wt{\cA}}$ are imposed.

\subsection{Evolution equation for $w=-2$}
To conclude our analysis, let us investigate the behaviour under symmetry transformations of the evolution equation relative to the covariant functional $\cT_{ab}$ of boost weight $w=-2$. Let us demonstrate that the following combination
\begin{equation}
\begin{aligned}
\cE_{\langle ab\rangle}^{\cT} &:= (\pa_v\cT_{ab})^\TF +\Bigl(2\mkell -\f{1}{2}\mthell \Bigl) \cT_{ab}- (\mr{D}_{\langle a}+5\mpi_{\langle a}) \cP_{b\rangle} +\f{3}{2}\mr{\sigma}^{(n)}_{ab}\cA +\f{3}{2}\mr{\wt{\sigma}}^{(n)}_{ab}\wt{\cA},
\label{ev_T}
\end{aligned}
\end{equation}
transforms without anomalies (on-shell). In particular, the trace-free part of $\pa_v\cT_{ab}$, i.e. $ (\pa_v\cT_{ab})^\TF = \pa_v \cT_{ab} -2 \cT_{c(a}\mr{\sigma}^{(\ell)c}_{b)}$, yields the following anomaly
\begin{equation}
\begin{aligned}
\Delta_\xi (\pa_v\cT_{ab})^\TF &= 4\pa_v(\cP_{\langle a} \pa_{b\rangle} \tau)- 4\mr{P}_{\langle a}\mr{\sigma}^{(\ell)}_{b\rangle c}\pa^c\tau -4 \mr{P}_{c} \mr{\sigma}^{(\ell)c}_{\langle a} \pa_{b\rangle}\tau. 
\end{aligned}
\end{equation}
The other anomaly terms read as follows
\begin{equation}
\begin{aligned}
&\Delta_\xi (\mr{D}_{\langle a} \cP_{b\rangle}) = \pa_v \cP_{\langle a}\pa_{b\rangle}\tau - \cP_{\langle a}\pa_{b\rangle} \dot{\tau} +\f{3}{2}\mr{D}_{\langle a}(\cA\pa_{b\rangle}\tau) +\f{3}{2}\mr{\wt{D}}_{\langle a}(\wt{\cA}\pa_{b\rangle}\tau) \\
&\qquad \qquad \qquad - 2\cP_c\mr{K}^{(\ell)c}_{\langle a} \pa_{b\rangle} \tau+ \cP_c \mr{\sigma}^{(\ell)}_{ab} \pa^c \tau,\\
&\Delta_\xi (\mpi_{\langle a} \cP_{b\rangle}) = \f{3}{2}\cA\mpi_{\langle a}\pa_{b\rangle}\tau + \f{3}{2}\wt{\cA}\mr{\wt{\pi}}_{\langle a}\pa_{b\rangle}\tau+\mr{\kappa}_{(\ell)}\cP_{\langle a}\pa_{b\rangle}\tau + \cP_{\langle a}\pa_{b \rangle}\dot{\tau}  - \cP_{\langle a} \mr{K}_{b\rangle}^{(\ell)c} \pa_c\tau,\\
&\Delta_\xi (\mr{\sigma}^{(n)}_{ab}\cA) = \cA\mr{D}_{\langle a}\pa_{b\rangle}\tau +2\cA\mpi_{\langle a} \pa_{b\rangle} \tau + 2\mr{\sigma}^{(n)}_{ab}\cJ^c\pa_c\tau.
\end{aligned}
\end{equation}
Using the relation introduced in the footnote \ref{note}, which gives
\begin{equation}
\Bigl(\mr{P}_c\mr{\sigma}^{(\ell)}_{ab} - \mr{P}_{\langle a}\mr{\sigma}^{(\ell)}_{b\rangle c}\Bigl)\pa^c\tau = \mr{P}_{c}\mr{\sigma}^{(\ell)c}_{\langle a} \pa_{b\rangle}\tau,
\end{equation}
and the fact that $\mr{\sigma}^{(n)}_{ab} \cJ^b= \mr{\wt{\sigma}}^{(n)}_{ab} \wt{\cJ}^b$, we obtain that the linear anomaly of $\eqref{ev_T}$ is
\begin{equation}
\begin{aligned}
\Delta_{\xi}\cE^\cT_{\langle ab\rangle} &=3\Bigl( \pa_v \cP_{\langle a} + \mr{\mu}_{(\ell)} \cP_{\langle a} - \cP_c \mr{\sigma}^{(\ell)c}_{\langle a}  -\f{1}{2}(\pa_{\langle a} +3\mpi_{\langle a})\cA -\f{1}{2}(\wt{\pa}_{\langle a} +3\mr{\wt{\pi}}_{\langle a})\wt{\cA}\\
&\qquad + 2\cJ^c \mr{\sigma}^{(n)}_{c \langle a} \Bigl) \pa_{b\rangle}\tau,
\end{aligned}
\end{equation}
that vanishes once the evolution equation of $\cP_a$ is imposed. Hence, the evolution equation \eqref{ev_T} transforms as follows 
\begin{equation}
\begin{aligned}
\var_{(\tau, Y)}\cE^\cT_{\langle ab\rangle}&=(\tau\pa_v +\Lie_Y -\dot{\tau})\cE_{\langle ab\rangle}^\cT +3\cE^\cP_{\langle a}\pa_{b\rangle}\tau.
\end{aligned}
\end{equation}
under the action of the near-horizon symmetry transformations.

\subsection{On-shell action}\label{sub_sec_3.4}
In the previous subsections, we derived the evolution equations of the (semi-) covariant functionals by looking for  combinations that transform homogeneously under the action of the full near-horizon symmetry group. In summary, we obtain the following set of evolution equations 
\begin{equation}
\begin{aligned}
& \pa_v \cJ_a -\Bigl(\mkell-\f{3}{2}\mthell\Bigl)\cJ_a - \mr{\sigma}^{(\ell)b}_{a} \cJ_b- (\mr{D}_b -\mpi_b)\cN^{\ b}_a=0,\\
&\pa_v\cA +\f{3}{2}\mthell\cA - (\mr{D}_a + \mpi_a)\cJ^a + \mr{\sigma}^{(n)}_{ab}\cN^{ab} =0,\\
&\pa_v\wt{\cA} +\f{3}{2}\mthell \wt{\cA} - (\mr{D}_a + \mpi_a)\wt{\cJ}^a + \mr{\sigma}^{(n)}_{ab}\wt{\cN}^{ab} =0,\\
&\pa_v\cP_a + \mr{\mu}_{(\ell)} \cP_a -\mr{\sigma}^{(\ell)b}_{a}\cP_b  -\f{1}{2}(\pa_a +3\mpi_a)\cA - \f{1}{2}(\wt{\pa}_a +3\mr{\wt{\pi}}_a)\wt{\cA} +2\mr{\sigma}^{(n)}_{ab}\cJ^b =0,\\
&(\pa_v\cT_{ab})^\TF +\Bigl(2\mkell -\f{1}{2}\mthell \Bigl) \cT_{ab} - (\mr{D}_{\langle a}+5\mpi_{\langle a}) \cP_{b\rangle} +\f{3}{2}\mr{\sigma}^{(n)}_{ab}\cA +\f{3}{2}\mr{\wt{\sigma}}^{(n)}_{ab}\wt{\cA} =0.
\end{aligned}
\label{ev_eq}
\end{equation}
As emphasized in the introduction, the main difference from the asymptotic case resides in the presence of genuine degrees of freedom on $\cH$, which implies
$\mthell\neq0, \ \mr{\sigma}^{(\ell)}_{ab}\neq0, \mkell\neq0, \mpi_a\neq0$.
In appendix \ref{NP_an}, we demonstrate that these equations can be obtained from the Newman-Penrose Bianchi identities, by exploiting the metric equations and the properties of duality symmetry.
From the behaviour of these evolution equations under the action of the near-horizon symmetry group, we can extract the following recursive pattern
\begin{equation}
\var_{(\tau,Y)}\cE^w = (\tau\pa_v + \Lie_Y +(w+1)\dot{\tau}) \cE^w + (2-w)\cE^{w+1}\pa\tau,
\label{pattern1}
\end{equation}
which is analogous to that found in \cite{Freidel:2021qpz, Geiller:2024bgf} for asymptotic null boundaries. In particular, imposing the constraints \eqref{ev_eq}, we obtain the on-shell action of the near-horizon symmetry group on the phase space variables. Thus, substituting the evolution equations \eqref{ev_eq} in the pattern \eqref{pattern1}, we have
\begin{equation}
\begin{aligned}
\var_{(\tau, Y)} \cJ_a &\heq \Bigl(\Lie_Y + \dot{\tau}+\tau(\mkell -2\mthell)\Bigl) \cJ_a 
+ \tau\mr{K}^{(\ell)b}_{a} \cJ_b + \tau(\mr{D}_b -\mpi_b)\cN^{\ b}_a 
+ \cN_a^{\ b}\pa_b\tau,\\
\var_{(\tau, Y)} \cA &\heq\Bigl(\Lie_Y- \f{3\tau}{2}\mthell\Bigl)\cA + \tau(\mr{D}_a + \mpi_a)\cJ^a - \tau\mr{\sigma}^{(n)}_{ab}\cN^{ab} +  2\cJ^a \pa_a \tau,\\
\var_{(\tau, Y)} \wt{\cA} &\heq \Bigl(\Lie_Y- \f{3\tau}{2}\mthell\Bigl)\wt{\cA}  + \tau(\mr{D}_a + \mpi_a)\wt{\cJ}^a - \tau\mr{\sigma}^{(n)}_{ab}\wt{\cN}^{ab}  +  2\wt{\cJ}^a \pa_a \tau,\\
\var_{(\tau, Y)}\cP_a &\heq (\Lie_Y - \dot{\tau} -\tau \mr{\mu}_{(\ell)}) \cP_a + \tau \mr{\sigma}^{(\ell)b}_{a}\cP_b  +\f{\tau}{2}(\pa_a +3\mpi_a)\cA + \f{\tau}{2}(\wt{\pa}_a +3\mr{\wt{\pi}}_a)\wt{\cA}\\
&\qquad -2\tau\mr{\sigma}^{(n)}_{ab}\cJ^b + \f{3}{2}\cA\pa_a\tau + \f{3}{2}\wt{\cA}\pa_a\tau,\\ 
\var_{(\tau, Y)}\cT_{ab}&\heq \Bigl[\Lie_Y -2\dot{\tau} -\tau\Bigl(2\mkell -\f{1}{2}\mthell \Bigl) \Bigl]\cT_{ab}+2\tau \cT_{c(a}\mr{\sigma}^{(\ell)c}_{b)} +\tau (\mr{D}_{\langle a}+5\mpi_{\langle a}) \cP_{b\rangle}\\
&\qquad  - \f{3\tau}{2}\mr{\sigma}^{(n)}_{ab}\cA   -\f{3\tau}{2}\mr{\wt{\sigma}}^{(n)}_{ab}\wt{\cA} + 4\cP_{\langle a} \pa_{b \rangle}\tau,
\end{aligned}
\end{equation}
and the evolution equations come from the action of super-translations on the covariant functionals and taking $\tau=1$, i.e. $\dot{\mathrm{Q}}_w = \var_{(1,0)} \mathrm{Q}_w$.

\section{Einstein-Cartan-Holst Lagrangian}\label{sec4}
As noted in \cite{Ruzziconi:2025fuy}, these covariant functionals of the metric appear in the sub-leading near-horizon Barnich-Brandt charges. Using the methodology of the covariant phase space \cite{Wald:1999wa,Barnich:2001jy, Donnelly:2016auv,Speranza:2017gxd,Harlow:2019yfa,Chandrasekaran:2020wwn,Freidel:2020xyx,Freidel:2021cjp}, the authors relate the sub-leading super-translation and $\cS$-diffeomorphism charges to the Weyl scalar $\Re\{\mr{\Psi}_2\}$ and $\mr{\Psi}_{3}$, respectively \footnote{To be precise, in our convention $\Psi_3$ and $\Psi_4$ have to be compared with the $\Psi_1$ and $\Psi_0$ of \cite{Ruzziconi:2025fuy}, respectively.}. It would be interesting to see if it is possible to recover the imaginary part of $\mr{\Psi}_2$ by adding topological terms to the Einstein-Hilbert Lagrangian. Following the analysis carried out at null infinity in \cite{Godazgar:2018dvh, Godazgar:2018qpq, Godazgar:2020kqd}, it was shown that the dual charges can be derived by adding the Holst term to the Einstein-Cartan Lagrangian.\\
The main goal of this section is to derive the covariant functional $\wt{\cA}$ (namely the imaginary part of $\mr{\Psi}_2$ as shown in appendix \ref{NP_an}) from the Holst pre-symplectic potential. For completeness, we also show how $\cA$ and $\cP_a$ appear in the Noether subleading charges, comparing our findings with those obtained in \cite{Ruzziconi:2025fuy}. In particular, in the computation of the sub-leading charges, we face a puzzle: our super-translation sub-leading charge does not coincide with the one obtained in \cite{Ruzziconi:2025fct}. This mismatch is resolved in section \ref{RZ_com} by adding a non-covariant boundary Lagrangian and using the formulae derived  in \cite{Freidel:2021cjp}. From the pre-symplectic potential ambiguity $\bd{\theta}\to \bd{\theta}' = \bd{\theta} + \var \bd{\ell}_b - \dext \bd{\vartheta}$, the Noether charges and fluxes in the two formulations are related through the following relations
\begin{equation}
    \cQ'_\xi = \cQ_\xi + \int_\cS (\iota_\xi \bd{\ell}_b -I_\xi\bd{\vartheta}), \qquad \cF'_\xi = \cF_\xi + \int_\cS (\var\iota_\xi \bd{\ell}_b -\var_\xi\bd{\vartheta}),
\label{q'f'}
\end{equation}
where $\bd{\ell}_b$ is the boundary Lagrangian and $\bd{\vartheta}$ is the corner pre-symplectic potential.

\subsection{Null tetrad}
Before delving into the construction of the sub-leading near-horizon charges, let us introduce the following frame fields
\begin{equation}
\hat{e}_0=\ell = \pa_\nu + V\pa_\rho -U^a\pa_a,\qquad \hat{e}_1 = n = \pa_\rho,\qquad
\hat{e}_i = E_i^a \pa_a, \label{null_tetrad}
\end{equation}
where
\begin{equation}
    E^a\pa_a= \sqrt{\f{q_{\theta\theta}}{2q}} \Bigl(\f{\sqrt{q} -iq_{\theta\phi}}{q_{\theta\theta}}\pa_\theta + i\pa_\phi \Bigl).
\end{equation}
The dual frame $\bd{e}^I$ can be computed straightforwardly, and its components read as follows
\begin{equation}
\bd{e}^0=\bd{n} = -\dext v, \qquad
\bd{e}^1 = \bd{\ell} =V\dext v - \dext \rho,\qquad
\bd{e}^i = E^i_a (\dext x^a + U^a \dext v),
\end{equation}
where
\begin{equation}
    E_a \dext x^a = \f{1}{\sqrt{2}} \sqrt{ q_{\theta \theta}}\ \dext \theta +\f{1}{\sqrt{2 q_{\theta\theta}}} (q_{\theta\phi} -i\sqrt{q})\ \dext \phi .
\end{equation}
In particular, using the results introduced in section \ref{review1}, the radial expansion of the complex dyad is
\begin{equation}
{e}^a_{\ i} = {e}^a_{(0) i} +\rho {e}^a_{(1) i}+ o(\rho),
\end{equation}
and specifically reads
\begin{equation}
\begin{aligned}
    {e}^a_{(0)i} = \mr{{E}}^a_{\ i}, \qquad
    {e}^a_{(1)i} = -\mr{K}_{c}^{(n)a} \mr{{E}}^{c}_{\ i},\qquad
    \cdots 
\end{aligned}
\end{equation}
and
\begin{equation}
\begin{aligned}
{m}_a^{(0)} = \mr{{E}}_a,\qquad
{m}_a^{(1)} = \mr{K}^{(n)c}_{a} \mr{{E}}_{c},\qquad
\cdots.
\end{aligned}
\end{equation}
Then, the near-horizon metric is defined as follows
\begin{equation}
g^{\mu\nu} = -n^{\mu}\ell^{\nu} - \ell^{\mu}n^\nu + m^{\mu}\bar{m}^{\nu} + \bar{m}^{\mu} m^\nu.
\end{equation}
In addition to the frame fields, in the tetrad formulation one needs to introduce the spin connection $\bd{\omega}^{IJ}$. Assuming a torsion-free spacetime, the components of the spin connection $\bd{\omega}^{IJ}$ are defined via the following relation
\begin{equation}
    \omega^{IJ}_\mu \dext x^\mu = e^{I}_\nu\nabla_\mu e^{\nu J}\ \dext  x^\mu.
\end{equation}
Explicitly, we obtain
\begin{equation}
\begin{aligned}
\omega^{01}_\mu \dext x^\mu&= \kappa_{(\ell)} \dext v + \pi_a(U^a\dext v + \dext x^a),\\ 
\omega^{1i}_\mu \dext x^\mu& = E^{ai}\Bigl(\pi_a \dext \rho - K^{(\ell)}_{ab}(\dext x^b + U^b \dext v) +(\pa_a V -V\pi_a) \dext v\Bigl), \\
\omega^{0i}_\mu \dext x^\mu &= -E^{ai} \Bigl(K^{(n)}_{ab}\dext x^b + (\pi_a + U^b K^{(n)}_{ab}) \dext v\Bigl),\\
\omega^{ij}_\mu \dext x^\mu &= E^{[i}_a \pa_\rho E^{j]a}  \dext \rho + \Bigl( E^{[i}_a \pa_v E^{j]a} + E^{a[i} E^{j]}_{\ b} D_a U^b\Bigl)\dext v +\omega^{ij}_{a} \dext x^a.\\
\end{aligned}
\end{equation}
The curvature tensor is defined as follows
\begin{equation}
    \bd{R}_{IJ} = \dext\bd{\omega}_{IJ} + \f{1}{2}[\bd{\omega},\bd{\omega}]_{IJ},
\end{equation}
and the Einstein-Cartan Lagrangian plus the Holst term is 
\begin{equation}
    \bold{L}_\ECH= \f{1}{2}\bd{\Sigma}_{IJ}\wedge \bd{R}^{IJ} ,
    \label{ECH_l}
\end{equation}
where
\begin{equation}
    \bd{\Sigma}_{IJ}=P_{IJKL}\ \bd{e}^K\wedge \bd{e}^L, \qquad \text{with}\quad P_{IJKL} = \f{1}{2}\epsilon_{IJKL}+\f{1}{\gamma}\eta_{I[K}\eta_{L]J}.
\end{equation}
Sometimes $\gamma$ is taken to be the Immirzi parameter, but here we consider it as a general parameter. The pre-symplectic potential is defined through the variation of the Lagrangian \eqref{ECH_l} and reads \cite{Freidel:2020svx}
\begin{equation}
    \bd{\theta}_\ECH [\bd{e}, \var\bd{\omega}] = \f{1}{2}\bd{\Sigma}_{IJ}\wedge \var\bd{\omega}^{IJ} = \f{1}{2} P_{IJKL} \bd{e}^K \wedge \bd{e}^L \wedge \var \bd{\omega}^{IJ},
    \label{th_ECH}
\end{equation}
and does not depend on $\var\bd{e}$. For convenience, we distinguish in the ECH pre-symplectic potential two contributions: the Einstein-Cartan pre-symplectic potential, denoted as follows
\begin{equation}
    {\bd{\theta}}_\EC= \f{1}{4} \epsilon_{IJKL} \bd{e}^K \wedge \bd{e}^L \wedge \var \bd{\omega}^{IJ} ,
\end{equation}
and the dual pre-symplectic potential, which is
\begin{equation}
    {\bd{\theta}}_\H= \f{1}{2\gamma} \eta_{I[K}\eta_{L]J} \bd{e}^K \wedge \bd{e}^L \wedge \var \bd{\omega}^{IJ}
    \label{H_th}
\end{equation}
and comes from the Holst term.

\subsection{Symmetries and charges}
The symmetry transformations of the tetrads read as follows
\begin{equation}
    \var_{(\xi,\lambda)} e^I_{\ \mu}= \Lie_\xi  e^I_{\ \mu} - \lambda^I_{\ J} e^J_{\ \mu}.
\end{equation}
The internal gauge transformations have to preserve the structure of the adapted metric, namely 
\begin{equation}
    e^0_{\ \rho}=0, \qquad e^0_{\ a}=0, \qquad e^1_{\ a}=0,
\end{equation}
from which we obtain 
\begin{equation}
\begin{aligned}
    \lambda^{0j} &=- E^{ja}\pa_a \xi^v,\\
    \lambda^{1j} &= E^{ja} (V\pa_a \xi^v -\pa_a \xi^\rho),\\
    \lambda^{10} &= \pa_\rho \xi^\rho,
\end{aligned}
\end{equation}
while $\lambda^{ij}$ remains unfixed. Using the results obtained in \cite{Freidel:2020xyx, Freidel:2020svx}, the Einstein-Cartan charges are defined as follows
\begin{equation}
    \cQ^{\EC}_{(\xi,\lambda)} =\int_\cS \dext^2 x \ \sqrt{q} (\iota_\xi \bd{\omega}^{10} +\lambda^{10}).
\end{equation}
Thus, by substituting the leading orders of the spin coefficient $\bd{\omega}^{10}$, we obtain 
\begin{equation}
\mr{\cQ}_T^{\EC} = -\int_\cS \dext^2 x \sqrt{\mr{q}}\ T\mkell,  \qquad
\mr{\cQ}^{\EC}_Y = -\int_\cS \dext^2 x \sqrt{\mr{q}}\  Y^a\mr{\pi}_a, \qquad
\mr{\cQ}_W^{\EC} = v\mr{\cQ}_{T=W}.
\end{equation}
The charge associated with internal gauge transformations yields the so-called internal Lorentz boost charge and reads
\begin{equation}
\mr{\cQ}_\lambda^{\EC} = -\int_\cS \dext^2 x \ \sqrt{\mr{q}}\ W.
\end{equation}
As shown in \cite{DeSimone:2025ouu}, the well-known Carrollian charges are obtained by adding a covariant Lagrangian
\begin{equation}
    \bd{\ell}_b = (\mkell + \mthell)\sqrt{\mr{q}}\ \dext v\ \dext^2 x,
\label{lb}
\end{equation}
and using the formula in \eqref{q'f'}, noticing that the corner potential is zero.

\noindent
These charges are well-known in literature, while the subleading structure of the near-horizon Einstein-Hilbert phase space has been analysed only in \cite{Ruzziconi:2025fuy}. In this section, we compare our sub-leading charges with the ones obtained in \cite{Ruzziconi:2025fuy} and analyse the sub-leading phase space coming from the presence of the Holst term. Knowing that $\sqrt{q}=(1+\rho\mtn )\sqrt{\mr{q}} + o(\rho) $, the sub-leading charge associated with the near-horizon super-translations is
\begin{equation}
\begin{aligned}
\cQ^{(1)}_\EC[\xi_T] &= -\int_\cS T\Bigl( \mr{\kappa}_{(\ell)}\mtn + \kappa^{(1)}_{(\ell)} +2\mpi^a\mpi_a -\mr{D}_a\mpi^a \Bigl)\ \sqrt{\mr{q}}\ \dext^2 x\\
&= \int_\cS T\Bigl(\mr{K}^{ab}_{(n)} \mr{K}^{(\ell)}_{ab} + \pa_v\mr{\theta}^{(n)} \Bigl)\ \sqrt{\mr{q}}\ \dext^2 x\\
&=- \int_\cS T\Bigl( \cA + \mkell\mtn - (\mr{D}_a + \mpi_a)\mpi^a
\Bigl)\ \sqrt{\mr{q}}\ \dext^2 x.
\end{aligned}
\end{equation}
The sub-leading super-translation charge does not coincide with that one obtained in \cite{Ruzziconi:2025fuy}, due to the presence of extra-terms. As we show in section \ref{RZ_com}, this inconsistency can be resolved by adding a non-covariant boundary Lagrangian. Now, let us compute the sub-leading charge associated with $\cS$-diffeomorphisms. Using the expansions furnished in section \ref{review1}, we obtain
\begin{equation}
\begin{aligned}
\cQ^{(1)}_\EC[\xi_Y] &= -\int_\cS Y^a \Bigl( \mr{D}_b\mr{\sigma}_a^{(n)b} - \f{1}{2}\mr{D}_a \mtn\Bigl)\ \sqrt{\mr{q}}\ \dext^2 x \\   
&= -\int_\cS Y^a \Bigl(\cP_a -\mpi_b \mr{\sigma}^{(n)b}_a +\f{1}{2}\mpi_a\mtn\Bigl)\ \sqrt{\mr{q}}\ \dext^2 x,   
\end{aligned}
\end{equation}
consistent with the sub-leading charge obtained in \cite{Ruzziconi:2025fuy}. In conclusion, the sub-leading Weyl charge is $\cQ^{(1)}_\EC[\xi_W]=v\cQ^{(1)}_\EC[\xi_{T=W}]$ and the sub-leading Lorentz charge is
\begin{equation}
\cQ^{(1)}_\EC[\lambda] = \int_\cS (-W\mtn + 2\mpi^a\pa_a (T+vW)) \ \sqrt{\mr{q}}\ \dext^2 x.  
\end{equation}

\subsection{Dual charges}
Now, we are interested in computing the Noether charges coming from the Holst contribution in the Einstein-Cartan-Holst pre-symplectic potential \eqref{th_ECH}. From the expression of the Holst pre-symplectic potential in \eqref{H_th}, the Holst Noether charge associated with a generic vector field $\xi$ reads as follows
\begin{equation}
\begin{aligned}
\cQ_\H[\xi] &\heq \f{1}{2\gamma}\int_\cS \eta_{I[K}\eta_{L]J} \bd{e}^K\wedge\bd{e}^L\wedge\iota_\xi\bd{\omega}^{IJ}\\
&= \f{1}{2\gamma}\int_\cS E_{i a} E_{jb}\Bigl[ \xi^v \Bigl( E^{[i}_c \pa_v E^{j]c} + E^{c[i} E^{j]}_{\ d} D_c U^d\Bigl) + \xi^\rho E^{[i}_c \pa_\rho E^{j]c} \\
&\qquad + \xi^c \omega^{ij}_c\Bigl] \dext x^a \wedge \dext x^b.
\end{aligned}
\label{QH_xi}
\end{equation}
Then, evaluating the equation \eqref{QH_xi} for $\xi=\xi_T$, the leading Holst charge associated with the super-translation symmetry is
\begin{equation}
\begin{aligned}
\mr{\cQ}_\H[\xi_T]&= \f{1}{2\gamma}\int_\cS T \mr{E}_{i a} \mr{E}_{jb}\ \Bigl( \mr{E}^{[i}_c \pa_v \mr{E}^{j]c} \Bigl)\ \dext x^a \wedge \dext x^b\\
&= -\f{1}{2\gamma}\int_\cS T \mr{K}^{(\ell)}_{ab} (\mr{\bar{E}}^a \mr{\bar{E}}^b - \mr{E}^a \mr{E}^b) \sqrt{\mr{q}}\  \dext^2 x,
\end{aligned}
\end{equation}
which contains the imaginary part of the spin-coefficient $\mr{\epsilon}_N$ (see appendix \ref{NP_an}). The sub-leading contribution is 
\begin{equation}
\begin{aligned}
\cQ^{(1)}_\H[\xi_T]&=  \f{1}{2\gamma} \int_\cS \epsilon_{ij} \Bigl[T \Bigl( \mr{K}^{(n)b}_{c} \mr{E}^{i}_b \pa_v \mr{E}^{jc} -\mr{E}^{i}_c \pa_v (\mr{K}^{(n)c}_{b} \mr{E}^{jb} ) + 2\mr{E}^{ci} \mr{E}^{j}_{\ d} \mr{D}_c \mpi^d\Bigl)\\
&\quad +  \mr{\omega}^{ij}_c\pa^c T \Bigl] \sqrt{\mr{q}}\  \dext^2 x + \f{1}{2\gamma}\int_\cS  \mtn \mr{q}_\H[\xi_T] \ \sqrt{\mr{q}}\ \dext^2x\\
&= \f{1}{2\gamma}\int_\cS \Bigl[-2T \Bigl(\wt{\cA} + \mr{\wt{\sigma}}^{(\ell)d}_{c} \mr{\sigma}^{(n)c}_{d}
\Bigl) + \epsilon_{ij} \mr{\omega}^{ij}_c \pa^{c} T +\mtn \mr{q}_\H[\xi_T]\Bigl]\sqrt{\mr{q}}\ \dext^2x.
\end{aligned}
\end{equation}
Similarly, one can extract the leading and subleading charge contributions coming from the $\cS$-diffeomorphisms and Weyl super-boosts, which yield
\begin{equation}
\begin{aligned}
\cQ_\H[\xi_Y] = \f{1}{2\gamma}\int_\cS \epsilon_{ij}Y^a\Bigl(\mr{\omega}^{ij}_a +\rho(\mtn\mr{\omega}^{ij}_a +\omega^{(1)ij}_a)\Bigl)\ \sqrt{\mr{q}}\ \dext^2 x,
\end{aligned}
\end{equation}
and
\begin{equation}
\cQ_\H[\xi_W] = v (\mr{\cQ}_\H[\xi_{T=W}] +\rho \cQ^{(1)}_\H[\xi_{T=W}]), 
\end{equation}
respectively. Hence, the inclusion of the Holst term in the Einstein-Cartan formulation reveals the existence of new, non-trivial near-horizon Noether charges.

\section{Non-covariant boundary Lagrangian} \label{RZ_com}
In the previous section, we faced an inconsistency: the sub-leading near-horizon super-translation charge we obtained does not coincide with the sub-leading super-translation Barnich-Brandt charge computed in \cite{Ruzziconi:2025fuy}. A similar problem was already encountered in the literature on null infinity, where the Noetherian charges constructed in \cite{Freidel:2021fxf} were not the same as the one considered in \cite{Barnich:2011mi}. In that case, the puzzle was resolved by a choice of a non-covariant boundary Lagrangian, providing a Noetherian construction of the charges obtained in \cite{Barnich:2011mi}. Therefore, we use the same reasoning. In our notation, the sub-leading super-translation charge (2.28) obtained in \cite{Ruzziconi:2025fuy} is the following
\begin{equation}
\begin{aligned}
\cQ^{RZ}_{\xi_T} &=  \int_\cS \Bigl((\pa_v + \mkell +\f{1}{2}\mthell)\mtn -\mr{D}_a\mpi^a -\mpi_a\mpi^a\Bigl)\sqrt{\mr{q}}\  \dext^2 x\\
&= - \int_\cS \Bigl(\cA + \mr{\sigma}^{(n)}_{ab} \mr{\sigma}_{(\ell)}^{ab}\Bigl)\sqrt{\mr{q}}\  \dext^2 x,
\end{aligned}
\label{RZ_charge}
\end{equation}
where we used the trivial equation in \eqref{E^c_c} in the following form
\begin{equation}
(\pa_v+\mkell)\mtn  - (\mr{D}_a + \mpi_a)\mpi^a +\f{1}{2}\mtn\mthell+ \mr{\sigma}^{(n)}_{ab} \mr{\sigma}_{(\ell)}^{ab}+\cA=0.
\end{equation}
Considering also the sub-leading contribution of the boundary Lagrangian required to transition from the Einstein-Cartan to the canonical formulation, a Noetherian construction of the sub-leading charge \eqref{RZ_charge} is obtained by adding the following boundary Lagrangian 
\begin{equation}
\begin{aligned}
\bd{\ell}^{(1)}_b &= \Bigl[\mr{\mu}_{\ell}\mtn + \kappa_{(\ell)}^{(1)} + \theta_{(\ell)}^{(1)} +\mpi_a\mpi_a\Bigl]\sqrt{\mr{q}}\ \dext v\ \dext^2 x\\
&= \Bigl[\mr{\mu}_{\ell}\mtn -\mr{K}^{(n)}_{ab} \mr{K}^{ab}_{(\ell)} - (\mr{D}_a+\mpi_a)\mpi^a \Bigl]\sqrt{\mr{q}}\ \dext v\ \dext^2 x.
\end{aligned}
\end{equation}
Indeed, using the radial expansions furnished in section \ref{review1}, the corner potential is still zero and 
from the relations in \eqref{q'f'} we obtain
\begin{equation}
\begin{aligned}
\cQ'_{(1)} [\xi_T]&= \cQ^{(1)}_\EC [\xi_T] + \int_\cS\iota_\xi\bd{\ell}^{(1)}_b\\
&=  \int_\cS \Bigl((\pa_v + \mr{\mu}_{(\ell)})\mtn -\mr{D}_a\mpi^a -\mpi_a\mpi^a\Bigl)\sqrt{\mr{q}}\  \dext^2 x\\
&=\cQ^{RZ}_{\xi_T}.
\end{aligned}
\end{equation}
The contribution of this boundary Lagrangian has to be taken into account to obtain the correct charge algebra. Indeed, as shown in \cite{Freidel:2021cjp} an extra term in the Barnich-Troessaert bracket \cite{Barnich:2011mi} occurs in order
to include the contributions from the Lagrangian and its anomaly. The contribution given from a non-covariant boundary Lagrangian enters in the bracket through the following term $\iota_\xi\Delta_\zeta\bd{\ell}_b - \xi \leftrightarrow \zeta $, where
\begin{equation}
\begin{aligned}
\Delta_\xi \bd{\ell}^{(1)}_b&= \Bigl[(\mr{D}_a +3\mpi_a )\mr{K}^{ab}_{(\ell)}\pa_b\tau - (2\pa_a +\mpi_a)\mkell\pa^a\tau +\f{1}{2}\mthell\mr{D}^2\tau \\
&\qquad - (\mr{D}_a+2\mpi_a) \pa^a\dot{\tau}\Bigl]\ \sqrt{\mr{q}}\ \dext v\ \dext^2 x.
\end{aligned}
\end{equation}

\section{Conclusion}
In this work, we have analysed the symmetry properties of the near-horizon covariant metric functionals that appear in the leading order of the Newman-Penrose Weyl scalars \cite{Ruzziconi:2025fuy}. Following the same strategy developed in \cite{Freidel:2021qpz}, we utilized the first (finite-distance) higher Bondi aspect $d_{\langle ab\rangle}$ to construct a functional of the metric that transforms semi-covariantly under the action of the near-horizon symmetry group and covariantly under boost transformations, with boost weight $w=-2$. The recursive pattern identified in the boost weight $w$ further allowed for the construction of remaining functionals with boost weights $w\in\{-1,0,1,2\}$. In particular, we emphasized the pivotal role of the duality symmetry in characterizing the phase space of a finite-distance null hypersurface and in evaluating the evolution equations of the covariant functionals. Specifically, we demonstrated that the dynamics of these boost-weighted functionals can be derived from symmetry arguments by identifying combinations of free data that transforms without anomalies under symmetry transformations.\\
As discussed throughout the work, the structure of the evolution equations in \eqref{ev_eq} is richer than the one found at null infinity, as the latter behaves as a non-expanding horizon \cite{Ashtekar:2021kqj, Ashtekar:2021wld, Ashtekar:2024bpi, Ashtekar:2024stm, Agrawal:2025fsv}. This richness naturally leads to some complications, such as the integration of the evolution equations under the no-radiation condition $\cN_{ab}=0$. In particular, the trace part of the non-radiation condition requires the null Raychaudhuri equation to hold, which can be solved non-perturbatively by trading the time variable for a dynamical field, called the dressing time which is canonically conjugate to the Raychaudhuri constraint \cite{Ciambelli:2023mir, Ciambelli:2024swv}. While this requires deeper investigation, it is already evident that the integration of \eqref{ev_eq} becomes straightforward when the horizon is non-expanding - a case we will explore in a forthcoming work.\\
As demonstrated in \cite{Ruzziconi:2025fuy}, the boost weighted $w=0$ and $w=-1$ functionals appear in the sub-leading contributions to the Barnich-Brandt charges, revealing a structural similarity between the sub-leading phase space at a finite distance and the leading radiative phase space at null infinity. However, the sub-leading super-translation charge in \cite{Ruzziconi:2025fuy} contains only the real part of the Weyl scalar $\Psi_2$ and it might be interesting to see if its imaginary part appears when adding topological terms to the Einstein-Cartan Lagrangian, in the same spirit of \cite{Godazgar:2020kqd, Oliveri:2020xls}. Our results show that this dual contribution emerges when computing the sub-leading Noether charge in the Einstein-Cartan-Holst formulation. Similarly to the asymptotic case, where the subleading BMS and dual charges are related to the Newman-Penrose charges, the derivation of the Holst/dual Noether charges presented in this work could be useful to capture the near-horizon Newman-Penrose (or Aretakis) charges \cite{Aretakis:2011ha, Aretakis:2011hc, Aretakis:2012ei}. \\
Finally, we also showed that the sub-leading super-translation charge identified in \cite{Ruzziconi:2025fuy} can be recovered by considering the charge contribution coming from adding a non-covariant boundary Lagrangian. The anomaly of this boundary Lagrangian enters in the charge algebra via the generalized Barnich-Troessaert bracket introduced in \cite{Freidel:2021cjp}, and therefore has to be taken into account in order to avoid field-dependent 2-cocycles.\\
To conclude, in addition to the future directions discussed throughout the manuscript, it might also be interesting to pursue the investigation of the near-horizon memory effects \cite{Donnay:2018ckb, Rahman:2019bmk} and to further clarify the relationship with these covariant functionals.

\appendix

\allowdisplaybreaks

\section{Noetherian and dual fluxes}\label{app_A}
In this appendix, we provide a derivation of the near-horizon Einstein-Cartan and Holst symplectic fluxes. Advocating the Noetherian split discussed in \cite{Freidel:2021cjp}, the Noetherian flux is defined as follows
\begin{equation}
    \cF_{(\xi,\lambda_L)} = \int_\cS(\iota_\xi\bd{\theta} + \bd{A}_\xi +\bd{q}_{\var\xi} + \bd{q}_{\var\lambda_L}),
\end{equation}
where the symplectic flux is
\begin{equation}
    \cF^\theta_\xi = \int_\cS\iota_\xi \bd{\theta} =  \int_\cS(\xi^r\theta^v - \xi^v\theta^r)\ \sqrt{\mr{q}}\ \dext^2 x.
\end{equation}
and $\bd{A}_\xi$ is the symplectic anomaly, whose exterior derivative is $\dext \bd{A}_\xi :=\Delta_\xi\bd{\theta} -\var\bd{a}_\xi +\bd{a}_{\var\xi}$ and $\bd{a}_\xi = \Delta_\xi\bd{L}$ is the Lagrangian anomaly.
However, the Einstein-Cartan formulation is anomaly-free when $\Lambda=0$, and therefore the contributions from the Lagrangian and symplectic anomaly vanish.  Using the expression of the Einstein-Cartan-Holst pre-symplectic potential in \eqref{th_ECH}, the symplectic flux then reads as follows
\begin{equation}
\begin{aligned}
\cF^\theta_{\ECH}[\xi] &= \int_\cS\iota_\xi\bd{\theta}_{\ECH}[\bd{e}, \var\bd{\omega}]\\
&= \f{1}{2}\int_\cS\iota_\xi\Bigl(P_{IJKL}\ \bd{e}^K\wedge\bd{e}^L\wedge\var\bd{\omega}^{IJ}\Bigl)\\
&= \f{1}{4}\int_\cS\iota_\xi \Bigl(\epsilon_{IJKL}\ \bd{e}^K\wedge\bd{e}^L\wedge\var\bd{\omega}^{IJ}\Bigl) +\f{1}{2\gamma}\int_\cS\iota_\xi \Bigl(\eta_{I[K}\eta_{L]J}\ \bd{e}^K\wedge\bd{e}^L\wedge\var\bd{\omega}^{IJ}\Bigl) ,
\end{aligned}
\end{equation}
where the first term in the last line is the Einstein-Cartan symplectic flux, while the second term is the Holst (or dual) symplectic flux. Let us compute these two contributions separately. The Einstein-Cartan flux relative to the generic vector field $\xi = \xi^v\pa_v + \xi^a\pa_a + \xi^\rho\pa_\rho$ is
\begin{equation}
\begin{aligned}
\cF_\xi^\theta &= \f{1}{4} \int_\cS \epsilon_{IJKL} \Bigl(2\iota_\xi\bd{e}^K \bd{e}^L \wedge \var \bd{\omega}^{IJ} +\bd{e}^K \wedge \bd{e}^L \iota_\xi\var \bd{\omega}^{IJ}\Bigl)\\
&= \f{1}{4} \epsilon_{IJKL} \int_\cS \xi^\mu \Bigl(e_{\ \mu}^K\ e_{\ a}^L \var \omega_{\ b}^{IJ}+\f{1}{2} e_{\ a}^K e^L_{\ b}\  \var \omega_{\ \mu}^{IJ}\Bigl)\ \dext x^a\wedge\dext x^b \\
&= -\f{1}{4} \epsilon_{IJKL} \int_\cS \xi^v \Bigl(e_{\ v}^K\ e_{\ a}^L \var \omega_{\ b}^{IJ}+\f{1}{2} e_{\ a}^K e^L_{\ b}\  \var \omega_{\ v}^{IJ}\Bigl)\ \dext x^a\wedge\dext x^b \\
&\quad +\f{1}{4}\int_\cS \xi^\rho \Bigl(-2 e_{\ \rho}^1 e_{\ a}^j \var \omega_{\ b}^{0i} \Bigl)\epsilon_{ij} \ \dext x^a\wedge\dext x^b.
\end{aligned}
\end{equation}
In particular, for $\xi = \xi^v\pa_v$ we have
\begin{equation}
\begin{aligned}
\cF_{\xi^v\pa_v}^\theta &= -\f{1}{4} \epsilon_{IJKL} \int_\cS \xi^v \Bigl(e_{\ v}^K\ e_{\ a}^L \var \omega_{\ b}^{IJ}+\f{1}{2} e_{\ a}^K e^L_{\ b}\  \var \omega_{\ v}^{IJ}\Bigl)\ \dext x^a\wedge\dext x^b \\
&= -\f{1}{4} \int_\cS \xi^v \Bigl(2 e_{\ v}^0\ e_{\ a}^j \var \omega_{\ b}^{1i} 
-2 e_{\ v}^1\ e_{\ a}^j \var \omega_{\ b}^{0i}  +2 e_{\ v}^i\ e_{\ a}^j \var \omega_{\ b}^{01} + e_{\ a}^i e^j_{\ b}\  \var \omega_{\ v}^{01}\Bigl)\ \epsilon_{ij}\  \dext x^a\wedge\dext x^b \\
&= \f{1}{4} \int_\cS \xi^v \Bigl(2 E_{\ a}^j \var \omega_{\ b}^{1i} +2V E_{\ a}^j \var \omega_{\ b}^{0i}  -2 U^c E_{\ c}^i\ E_{\ a}^j \var \omega_{\ b}^{01} - E_{\ a}^i E^j_{\ b}\  \var \omega_{\ v}^{01}\Bigl)\ \epsilon_{ij}\  \dext x^a\wedge\dext x^b \\
&= \f{1}{2} \int_\cS \xi^v \Bigl(2  \var \omega_{\ b}^{1i} +2V  \var \omega_{\ b}^{0i}  -2 U^c E_{\ c}^i \var \omega_{\ b}^{01} \Bigl)\ E^{\ b}_i \sqrt{q}\ \dext^2 x - \int_\cS \xi^v \var \omega_{\ v}^{01}\ \sqrt{q}\  \dext^2 x\\
&= -\f{1}{2} \int_\cS \xi^v \Bigl(2  \var (E^{ic}K^{(\ell)}_{cb}) +2V  \var (E^{ic}K^{(n)}_{cb})  +2 U^c E_{\ c}^i \var \pi_{b} \Bigl)\ E^{\ b}_i \sqrt{q}\ \dext^2 x\\
&\quad - \int_\cS \xi^v \var (\kappa_{(\ell)} + \pi_cU^c)\ \sqrt{q}\  \dext^2 x\\
&= -\f{1}{2} \int_\cS \xi^v \Bigl(2  \var\thell +2V  \var\theta^{(n)} - (K^{(\ell)}_{cb} + VK^{(n)}_{cb})\var q^{bc} -2 U^b\var \pi_{b} \Bigl)\  \sqrt{q}\ \dext^2 x\\
&\quad - \int_\cS \xi^v \var (\kappa_{(\ell)} + \pi_cU^c)\ \sqrt{q}\  \dext^2 x,
\end{aligned}
\end{equation}
and for $\xi=\xi^\rho\pa_\rho$, we obtain
\begin{equation}
\begin{aligned}
\cF_{\xi^\rho\pa_\rho}^\theta &= -\f{1}{2}\int_\cS \xi^\rho  E_{\ a}^j \var (E^{ci}K^{(n)}_{cb}) \epsilon_{ij} \ \dext x^a\wedge\dext x^b\\
&= -\int_\cS \xi^\rho  \var (E^{ci} K^{(n)}_{cb})E^b_{\ i}\ \sqrt{q}\ \dext^2 x\\
&= -\int_\cS \xi^\rho \Bigl(\var\theta^{(n)} -\f{1}{2} K^{(n)}_{cb}\var q^{cb} \Bigl)\ \sqrt{q}\ \dext^2 x.\\
\end{aligned}
\end{equation}

\noindent
Let us now focus on the Holst contribution to the symplectic flux. From the expression of the Holst pre-symplectic potential in \eqref{H_th}, we have
\begin{equation}
\begin{aligned}
\cF_\H^\theta [\xi] &= \f{1}{2\gamma} \int_\cS \eta_{I[K}\eta_{L]J}\Bigl(2\iota_\xi\bd{e}^K  \bd{e}^L \wedge \var \bd{\omega}^{IJ} +\bd{e}^K \wedge \bd{e}^L \iota_\xi\var \bd{\omega}^{IJ} \Bigl)\\
&= \f{1}{4\gamma} \int_\cS \eta_{I[K}\eta_{L]J} \xi^\mu \Bigl(2 e_{\ \mu}^K  e^L_{\ a} \var \omega^{IJ}_{\ b} + e^K_{\ a} e_{\ b}^L \var \omega_{\ \mu}^{IJ} \Bigl)\ \dext x^a \wedge \dext x^b\\
&= \f{1}{4\gamma} \int_\cS \xi^\mu \Bigl(2 e_{0\mu}  e_{j a} \var \omega^{0j}_{\ b} +2 e_{1\mu}  e_{j a} \var \omega^{1j}_{\ b} +2 e_{i\mu}  e_{j a} \var \omega^{ij}_{\ b}  + e_{ia} e_{jb} \var \omega_{\ \mu}^{ij} \Bigl)\ \dext x^a \wedge \dext x^b.\\
\end{aligned}
\end{equation}
Then, we have the following two contributions:
\begin{equation}
\begin{aligned}
\cF_\H^\theta [\xi^v\pa_v]&=- \f{1}{4\gamma} \int_\cS \xi^v \Bigl[-2VE_{j a} \var \omega^{0j}_{\ b} +2 E_{j a} \var \omega^{1j}_{\ b} +2 U^cE_{ic}  e_{j a} \var \omega^{ij}_{\ b}\\
&\quad+ e_{ia} e_{jb} \var ( E^{[i}_c \pa_v E^{j]c} + E^{c[i} E^{j]}_{\ d} D_c U^d) \Bigl]\ \dext x^a \wedge \dext x^b\\
&=- \f{1}{4\gamma} \int_\cS \xi^v \Bigl(2VE_{j a} \var (E^{cj}{K}^{(n)}_{cb}) -2 E_{j a} \var (E^{cj}{K}^{(\ell)}_{cb}) +2 U^cE_{ic}  E_{j a} \var \omega^{ij}_{\ b}\\
&\quad+ E_{ia} E_{jb} \var ( E^{[i}_c \pa_v E^{j]c} + E^{c[i} E^{j]}_{\ d} D_c U^d) \Bigl)\ \dext x^a \wedge \dext x^b\\
&= -\f{1}{4\gamma} \int_\cS \xi^v \Bigl(2q_{\ a}^c(V\var K^{(n)}_{cb} -\var {K}^{(\ell)}_{cb}) 
+ (VK^{(n)}_{cb} - {K}^{(\ell)}_{cb})\var q^c_{\ a}  +2 U^cE_{ic}  E_{j a} \var \omega^{ij}_{\ b}\\
&\quad+ E_{ia} E_{jb} \var ( E^{[i}_c \pa_v E^{j]c} + E^{c[i} E^{j]}_{\ d} D_c U^d) \Bigl)\ \dext x^a \wedge \dext x^b,
\end{aligned}
\end{equation}
and
\begin{equation}
\begin{aligned}
\cF_\H^\theta [\xi^\rho\pa_\rho] &= \f{1}{4\gamma} \int_\cS \xi^\rho \Bigl(2e_{j a} \var \omega^{0j}_{\ b} + e_{ia} e_{jb} \var (E^{[i}_c \pa_\rho E^{j]c}) \Bigl)\ \dext x^a \wedge \dext x^b\\
 &= \f{1}{4\gamma} \int_\cS \xi^\rho \Bigl(-2 E_{j a} \var (E^{cj}K^{(n)}_{cb}) + E_{ia} E_{jb} \var (E^{[i}_c \pa_\rho E^{j]c}) \Bigl)\ \dext x^a \wedge \dext x^b\\
 &= -\f{1}{4\gamma} \int_\cS \xi^\rho \Bigl(2q^c_{\ a}
\var K^{(n)}_{cb} + K^{(n)}_{cb} \var q^c_{\ a}  - E_{ia} E_{jb} \var (E^{[i}_c \pa_\rho E^{j]c}) \Bigl)\ \dext x^a \wedge \dext x^b.
\end{aligned}
\end{equation}

\section{Newman-Penrose analysis}\label{NP_an}
Now, to be consistent with our treatment, in this appendix we present a Newman-Penrose analysis of the near-horizon geometry, comparing our results with the ones obtained in \cite{Ruzziconi:2025fuy} \footnote{Notice that in this work we have used a different notation. In particular, the roles of $\ell$ and $n$ must be switched in order to compare our results with the ones in \cite{Ruzziconi:2025fuy}, i.e., our $\Psi_4$ is equal to $\Psi_0^{RZ}$ in \cite{Ruzziconi:2025fuy}.}. 
The near-horizon geometry \eqref{metric} can be described by choosing the null frame ${e}^\mu_I = ({\ell}^\mu, {n}^\mu, {m}^\mu, {\bar{m}}^\mu)$ introduced in \eqref{null_tetrad}. The spin coefficients are computed using the following formula
\begin{equation}
    \gamma_{IJK} = e^{\ \mu}_I e^{\ \nu}_K \nabla_\nu e_{J\ \mu}
\end{equation}
with $\gamma_{IJK} = -\gamma_{JIK}$. Explicitly, we use the following convention for the spin coefficients
\begin{equation}
\begin{aligned}
\kappa &= \gamma_{131},\qquad \varrho&=\gamma_{134}, \qquad \epsilon &=\frac{1}{2}(\gamma_{121} -\gamma_{341}), \\
\sigma &= \gamma_{133},\qquad \mu&=\gamma_{423}, \qquad \gamma &=\frac{1}{2}(\gamma_{122} -\gamma_{342}), \\
\lambda &= \gamma_{424},\qquad \tau&=\gamma_{132}, \qquad \alpha &=\frac{1}{2}(\gamma_{124}-\gamma_{344}), \\
\nu &= \gamma_{422},\qquad \pi&=\gamma_{421}, \qquad \beta &=\frac{1}{2}(\gamma_{123}-\gamma_{343}). \\
\end{aligned}
\end{equation}
The complex conjugate of any spin coefficient is obtained  by replacing the index 3 by the index 4 whenever it occurs, and vice versa.\\
The leading order of the boundary metric and the volume form on $\cS$ introduced in subsection \ref{area_dual} can be written in terms of the complex dyad as follows
\begin{equation}
    \mr{q}^{ab} = \mr{E}^a\mr{\bar{E}}^b +\mr{\bar{E}}^a \mr{E}^b, \qquad i\varepsilon^{ab} = \mr{E}^a\mr{\bar{E}}^b -\mr{\bar{E}}^a \mr{E}^b,
\end{equation}
and by defining the mixed tensor $i\varepsilon_a^{\ b} :=i\varepsilon_{ac}q^{cb} =\mr{E}_a\mr{\bar{E}}^b  -\mr{\bar{E}}_a \mr{E}^b$, we have
\begin{equation}
    \varepsilon_a^{\ b}\varepsilon^{\ a}_b =-2, \qquad \varepsilon_a^{\ b} \mr{E}^a = i\mr{E}^b, \qquad \varepsilon_a^{\ b} \mr{\bar{E}}^a = -i\mr{\bar{E}}^b.
\end{equation}
The directional derivative along the vector fields $e^\mu_I$ can be compactly written as follows
\begin{equation}
\Delta := n^\mu\pa_\mu, \qquad D := \ell^\mu\pa_\mu, \qquad \delta := m^\mu\pa_\mu, \qquad \bar{\delta} := \bar{m}^\mu\pa_\mu.
\end{equation}
Before delving in the computation of the spin coefficients and Weyl scalars, it is fundamental to introduce the notion of weighted operators.

\subsection*{Weighted operators}
The null tetrad defined previously can be written in terms of a pair of spinor field $o^A$ and $\iota^A$ as follows
\begin{equation}
    \ell^\mu = o^A\bar{o}^{A'}, \qquad n^\mu = \iota^A\bar{\iota}^{A'}, \qquad m^\mu = o^A\bar{\iota}^{A'}, \qquad \bar{m}^\mu = \iota^A\bar{o}^{A'}
\end{equation}
with $A=0,1$ and we set $o_A\iota^A=1$. The boost and spin transformations, respectively
\begin{equation}
    \ell^\mu \to e^{\lambda_L}\ell^\mu, \qquad n^\mu \to e^{-\lambda_L}n^\mu,
\end{equation}
and
\begin{equation}
    m^\mu \to e^{i\chi}m^\mu, \qquad \bar{m}^\mu \to e^{-i\chi}\bar{m}^\mu,
\end{equation}
reads as follows in the spin frame ($o^A, \iota^A$),
\begin{equation}
    o^A\to t o^A,\qquad \iota^A\to t^{-1}\iota^A,\label{oa_tr}
\end{equation}
where $t$ is a nowhere vanishing complex scalar field. Therefore, a quantity $\eta$ is of type $\{p,q\}$ if transforms as follows 
\begin{equation}
    \eta \to t^{p}\bar{t}^{q}\eta
\end{equation}
under \eqref{oa_tr}. Equivalently, we say that $\eta$ has spin weight $s=(p-q)/2$ and boost weight $w=(p+q)/2$. At this point, one introduce the following weighted operators
\begin{equation}
\begin{aligned}
\pth \eta &= (D - p\epsilon -q\bar{\epsilon} )\eta,\qquad  \pth' \eta = (\Delta -p\gamma -q\bar{\gamma})\eta, \\
\eth \eta &= (\var -p\beta -q\bar{\alpha})\eta, \qquad
\bar{\eth} \eta = (\bar{\var} -p\alpha-q\bar{\beta} )\eta,\\
\end{aligned}
\label{sw_oper}
\end{equation}
with $\{p,q\}$-weights 
\begin{equation}
    \pth=\{1,1\}, \qquad \pth'=\{-1,-1\}, \qquad \eth=\{1,-1\}, \qquad \bar{\eth} = \{-1,1\}.
\end{equation}
In addition to the transformations in \eqref{oa_tr}, one has to consider the conformal rescaling
\begin{equation}
    o^A\to\Omega^{\omega_0}o^A, \qquad \iota^A\to\Omega^{\omega_1}\iota^A,
    \label{weyl_oa}
\end{equation}
where $\Omega$ is a definite-positive scalar field and $\omega_{0,1}$ are the Weyl weights of the spinorial fields. Then, we say that a quantity $\eta$ has Weyl weight $\omega$ if transforms as follows
\begin{equation}
    \eta\to\Omega^\omega \eta
\end{equation}
under \eqref{weyl_oa}. As previously, one introduces the following Weyl-weighted operators
\begin{equation}
\begin{aligned}
\pth_{\scr{C}}  &= \pth + [\omega + (p+q)\omega_1]\rho,\qquad & \pth'_{\scr{C}} = \pth' -[\omega - (p+q)\omega_0]\mu, \\
\eth_{\scr{C}}  &= \eth + [\omega +p\omega_1 -q\omega_0]\tau, \qquad&
\bar{\eth}_{\scr{C}}  = \bar{\eth} - [\omega -p\omega_0 +q\omega_1]\pi,
\end{aligned}
\end{equation}
whose $\{p,q,\omega\}$-weights are
\begin{equation}
\begin{aligned}
    \pth_{\scr{C}}&=\{1,1,2\omega_0\}, \qquad& \pth'_{\scr{C}}=\{-1,-1, 2\omega_1\}, \\
    \eth_{\scr{C}}&=\{1,-1, \omega_0+\omega_1\}, \qquad& \bar{\eth}_{\scr{C}} = \{-1,1,\omega_0+\omega_1\}.
\end{aligned}
\end{equation}
The Bianchi identities are covariant under the transformations in \eqref{oa_tr} and \eqref{weyl_oa}, and following the same conventions of \cite{Penrose:1985bww}, the Weyl scalars have the following $\{p, q,\omega\}$-weights
\begin{equation}
\begin{aligned}
\Psi_0=\{4,0,4\omega_0-1\},& \qquad \Psi_1=\{2,0,3\omega_0+\omega_1-1 \}, \qquad \Psi_2=\{0,0, 2\omega_0+2\omega_1-1\},\\
&\Psi_3=\{-2,0, \omega_0+3\omega_1-1\}, \qquad \Psi_4=\{-4,0,4\omega_1-1\}.
\end{aligned}
\end{equation}

\subsection*{Spin coefficients}
As emphasized in \cite{Geiller:2024bgf}, many subtleties occur when writing solutions obtained in the metric formulation into the Newman-Penrose formalism. Therefore, in order to avoid these mismatches, the metric-NP dictionary worked out in \cite{Geiller:2024bgf} must be carefully followed.
Firstly, we have to require $\ell$ and ${m}$ to be parallel propagated along $n^\mu$, i.e.,
\begin{equation}
    \Delta n = 0, \qquad \Delta \ell=0, \qquad \Delta m=0,
\end{equation}
that means $\gamma= \tau = \nu=0$. However, using the null tetrad introduced in \eqref{null_tetrad}, we have \begin{equation*}
\begin{aligned}
{\kappa} &= \pa_a V\ {m}^a, \qquad\qquad& {\gamma} &= \f{1}{8}\pa_\rho q_{ab} (\bar{{m}}^a \bar{{m}}^b- {m}^a {m}^b )\\ 
{\tau} &= \f{1}{2} q_{ab}\pa_\rho U^a {m}^b , \qquad\qquad&  {\lambda} &= K^{(n)}_{ab} \bar{{m}}^a \bar{{m}}^b\\
{\pi} &=\f{1}{2} q_{ab}\pa_\rho U^a {{\bar{m}}}^b , \qquad\qquad&  {\epsilon} &= \f{1}{2} \kappa_{(\ell)} - \f{1}{2}{m}^a \nabla_\ell \bar{{m}}_a\\
{\nu} &=0, \qquad\qquad& {\varrho} &= -\f{1}{2}\theta^{(\ell)} \\
{\mu}&= \f{1}{2} \theta^{(n)}, \qquad\qquad& {\alpha} &=\f{1}{4} q_{ab}\pa_\rho U^a {\bar{m}}^b - \f{1}{2}(D_b {\bar{m}}_a) {\bar{m}}^b {m}^a \\
{\sigma}&= -K^{(\ell)}_{ab} {m}^a {m}^b, \qquad\qquad& {\beta} &=\f{1}{4} q_{ab}\pa_\rho U^a {m}^b + \f{1}{2}(D_b {m}_a) {m}^b {\bar{m}}^a\\
\end{aligned}
\end{equation*}
and see that $\Re{\gamma}=0$ and $\Im{\mu}=0$, so the twist is zero. Therefore, we have to perform Lorentz transformations in order to set to zero the spin coefficients ${\tau}$ and ${\gamma}$. Recall that we have three classes of rotations. Given $a$ and $b$ complex functions and $A$ and $\chi$ real functions, we have
\begin{itemize}
\item[(a)] \emph{rotations of class I} which leave ${\ell}$ unchanged:
\begin{equation}
\begin{aligned}
{\ell}\to {\ell}, &\qquad {m}\to {m} +a{\ell}, \qquad {\bar{m}}\to {\bar{m}}+\bar{a}  {\ell}\\
&{n}\to{n} +\bar{a}   {m} +a {\bar{m}} +a\bar{a}  {\ell};
\end{aligned}
\label{I}
\end{equation}
\item[(b)] \emph{rotations of class II} which leave ${n}$ unchanged:
\begin{equation}
\begin{aligned}
{n}\to {n}, &\qquad {m}\to {m} +b{n}, \qquad {\bar{m}}\to {\bar{m}}+\bar{b}  {n}\\
&{\ell}\to{\ell} +\bar{b}   {m} +b {\bar{m}} +b\bar{b}  {n};
\end{aligned}
\label{II}
\end{equation}
\item[(c)] \emph{rotations of class III} which leave the direction of ${\ell}$ and ${n}$ unchanged:
\begin{equation}
{\ell}\to A^{-1}{\ell}, \qquad {n}\to A{n}, \qquad {m}\to e^{i\chi}{m}, \qquad {\bar{m}}\to e^{-i\chi}{\bar{m}}.
\end{equation}
\label{III}
\end{itemize}

\vspace{-0.6cm}
\noindent
We want to preserve the direction of $n$, therefore we apply a combination of class II and III transformations, yielding
\begin{equation}
\begin{aligned}
    {n}\to A {n}, \qquad {\ell}\to A^{-1} {\ell} +\bar{b}  e^{i\chi}  {m} +be^{-i\chi}{\bar{m}} + A b\bar{b} {n}\\
    {m} \to e^{i\chi}{m} + Ab {n}, \qquad     {\bar{m}} \to e^{-i\chi}{\bar{m}} + A\bar{b}n.
\end{aligned}
\label{II_III}
\end{equation}
Then, the spin coefficients in exam transform as follows
\begin{equation}
    {\tau} \to e^{i\chi}{\tau} , \qquad {\gamma} \to A{\gamma} -\f{1}{2}{n}^\nu (\pa_\nu A -iA\pa_\nu\chi), \qquad {\nu}\to e^{-i\chi}A^2{\nu}, \qquad {\mu} \to A {\mu}
\end{equation}
 under the III class, and
\begin{equation}
    {\tau} \to {\tau}+2b{\gamma} -{n}^\nu\pa_\nu b +b^2{\nu},  \qquad {\nu}\to {\nu},
    \qquad {\gamma} \to {\gamma} + b{\nu},  \qquad {\mu} \to {\mu} + b{\nu}   
\end{equation}
under the II class. Then, by identifying the new spin-coefficients with the subscript $_N$, we have
\begin{equation}
    \begin{aligned}
        \tau_N &= e^{i\chi} {\tau} + 2 b \Bigl(A{\gamma} -\f{1}{2}{n}^\nu (\pa_\nu A -iA\pa_\nu\chi)\Bigl) -A{n}^\nu\pa_\nu b=0, \\
        \gamma_N &= A{\gamma} -\f{1}{2}{n}^\nu (\pa_\nu A -iA\pa_\nu\chi)  =0,\\
        \mu_N &= A{\mu}.
    \end{aligned}
\end{equation}
From the last equation, we set $A=1$ and
\begin{equation}
    e^{i\chi}{\tau} - \pa_\rho b=0 \quad \Rightarrow \quad b = \int d\rho\ e^{i\chi}{\tau}.
\end{equation}
Finally, one obtains
\begin{equation}
    A=1, \qquad \chi = 2i\int d\rho\ \Im{{\gamma}}, \qquad b = \int d\rho\ e^{i\chi}{\tau}.
\end{equation}
Now, substituting the expressions of ${\tau}$ and ${\gamma}$ in the integrals above, we get
\begin{equation}
\begin{aligned}
\chi &= \f{i}{4} \int d\rho\ \pa_\rho q_{ab} ( {\bar{m}}^a{\bar{m}}^b - {m}^a{m}^b),\\
b&=\f{1}{2}\int d\rho\ e^{i\chi} q_{ab}\pa_\rho U^a {m}^b .
\end{aligned}
\end{equation}
From now on, we use the transformed tetrad $(\ell^\mu_{N},n^\mu_{N},m^\mu_{N},\bar{m}^\mu_{N})$ to compute the Newman-Penrose scalars. The new tetrad then reads
\begin{equation}
\begin{aligned}
\ell_N &= \pa_v + \rho\mr{\kappa}_{(\ell)} \pa_\rho - \rho(\mr{\bar{\pi}}\mr{\bar{E}}^a + \mpi \mr{E}^a)\pa_a + o(\rho),\\
 m_N &=  \rho\mr{\bar{\pi}}\pa_\rho + \Bigl(\mr{E}^a -\rho( \mr{\bar{\lambda}}\mr{\bar{E}}^a +\mr{\mu}\mr{E}^a) \Bigl)\pa_a+ o(\rho).
\end{aligned}
\end{equation}
In particular, except for the $\kappa_N$ (and obviously for $\gamma_N$ and $\tau_N$ which now vanish), the leading order of the spin coefficients and Weyl scalars does not change under Lorentz transformations of class II and III, i.e., $\mr{\lambda}_N=\mr{\lambda}$ and $\mr{\Psi}_{N4} = \mr{\Psi}_{4}$. Then, the new spin coefficients read as follows
\begin{equation}
\begin{aligned}
    \kappa_N &= \rho (\mr{D}_a \mr{\kappa}_{(\ell)} -\pa_v\mpi_a )\mr{E}^a + o(\rho), \\
    \pi_N &= \mr{\pi}_a \mr{\bar{E}}^a -\rho\Bigl(\mpi\mr{\mu} + \mr{\bar{\pi}}\mr{\lambda} +\mr{\Psi}_3\Bigl) +o(\rho), \\
    \mu_N &= \f{1}{2}\mtn -\rho(\mr{\mu}^2 + \mr{\lambda}\mr{\bar{\lambda}})+o(\rho), \\
    \lambda_N &= \mr{K}^{(n)}_{ab} \mr{\bar{E}}^a \mr{\bar{E}}^b - \rho\Bigl( 2\mr{\lambda}\mr{\mu} + \mr{\Psi}_4\Bigl)+o(\rho),\\
    \epsilon_N &= \f{1}{2}\Bigl(\mr{\kappa}_{(\ell)}  +\f{1}{2}\mr{K}_{ab}^{(\ell)}(\bar{E}^a\bar{E}^b - E^a E^b) \Bigl) - \rho( \mr{\bar{\pi}}\mr{\alpha} +\mpi\mr{\beta} + \mr{\Psi}_2) +o(\rho), \\
    \varrho_N &= -\f{1}{2}\mthell  - \rho(\mr{\mu}\mr{\varrho} +\mr{\sigma}\mr{\lambda} + \mr{\Psi}_2) + o(\rho),\\
    \alpha_N &= \f{1}{2}(\mpi_a -  \mr{D}_a) \mr{\bar{E}}^a -\rho( \mr{\beta}\mr{\lambda} + \mr{\mu}\mr{\alpha} + \mr{\Psi}_3) +o(\rho),  \\
    \beta_N &= \f{1}{2}(\mpi_a + \mr{D}_a) \mr{E}^a -\rho(\mr{\mu} \mr{\beta} + \mr{\alpha}\mr{\bar{\lambda}}) + o(\rho), \\
    \sigma_N &= -\mr{K}^{(\ell)}_{ab} \mr{E}^a \mr{E}^b -\rho(\mr{\mu}\mr{\sigma} + \mr{\varrho}\mr{\bar{\lambda}}) + o(\rho). 
\end{aligned}
\label{list_sc}
\end{equation}
Under the boost and spin transformations discussed in the previous subsection (see the book \cite{Penrose:1985bww} for details), the spin coefficients which admit well-definite $\{s, w\}$-weights are $\kappa, \nu, \rho, \mu, \sigma, \lambda, \tau, \pi$, while the remaining spin coefficients admit extra terms under these transformations and are used to define the weighted operators $\pth, \pth', \eth, \bar{\eth}$ in \eqref{sw_oper}. For convenience, we always work with the leading order of these weighted operator, i.e. $\eth=\mr{\eth}$. In particular, for an arbitrary quantity $X_{(s,w)}$ of spin weight $s$ and boost weight $w$, we have 
\begin{equation}
    \mr{D}X_{(s,w)} = (\eth  + w\mr{\bar{\pi}} )X_{(s,w)},\qquad \text{and} \qquad \mr{\bar{D}}X_{(s,w)} = (\bar{\eth} + w\mpi) X_{(s,w)},
\label{DD}
\end{equation}
where $\mr{D}= \mr{E}^a\mr{D}_a$ and $\mr{\bar{D}}= \mr{\bar{E}}^a\mr{D}_a$.

\subsection*{Weyl scalars}
The five complex Weyl scalars are defined as follows
\begin{equation}
\begin{aligned}
    \Psi_0 = -C_{abcd}\ell^a m^b & \ell^c m^d, \qquad     \Psi_1 = -C_{abcd}\ell^a m^b\ell^c n^d, \qquad     \Psi_2 = -C_{abcd}\ell^a m^b\bar{m}^c n^d,\\
    &\Psi_3 = -C_{abcd}\ell^a n^b\bar{m}^c n^d, \qquad
    \Psi_4 = -C_{abcd}\bar{m}^a n^b\bar{m}^c n^d.
\end{aligned}
\end{equation}
Having computed the NP spin coefficients, the Weyl scalars can be evaluated by using the NP angular equations, the radial equation of $\lambda_N$ and the evolution equation of $\sigma_N$. In particular, their subleading terms can be evaluated through the radial Bianchi identities 
\begin{equation}
\begin{aligned}
\Delta\Psi_{N0} - \delta\Psi_{N1} &= -\mu_N\Psi_{N0} -2\beta_N\Psi_{N1} +3\sigma_N\Psi_{N2},\\
\Delta\Psi_{N1} - \delta\Psi_{N2} &= -2\mu_N\Psi_{N1} +2\sigma\Psi_3,\\
\Delta\Psi_{N2} - \delta\Psi_{N3} &= -3\mu_N\Psi_{N2} +2\beta_N\Psi_{N3} + \sigma_N\Psi_{N4},\\
\Delta\Psi_{N3} - \delta\Psi_{N4} &= -4\mu_N\Psi_{N3} +4\beta_N\Psi_{N4}\\
\end{aligned}
\label{radBI}
\end{equation}
once $\Psi_{N4}$ is specified on $\cH$. Indeed, $\Psi_{N4}$ is a free data on $\cH$ and its value is fundamental to access to the sub-leading orders of the other Weyl scalars. Let us see how to compute its leading and sub-leading expression.

\subsubsection*{Computation of $\Psi_4$}
Using the metric equation (relative to the null tetrad in \eqref{null_tetrad})
\begin{equation}
    \pa_\rho m^a = -\bar{\lambda}\bar{m}^a - (\mu-\gamma+\bar{\gamma})m^a
\end{equation}
the Weyl scalar $\Psi_4$ relative to the tetrad in \eqref{null_tetrad} is
\begin{equation}
\begin{aligned}
\Psi_4 &= -\pa_\rho\lambda -(2\mu + 3\gamma - \bar{\gamma})\lambda \\
&= -(\pa_\rho K^{(n)}_{ab}) \bar{m}^a\bar{m}^b - K^{(n)}_{ab} \pa_\rho( \bar{m}^a\bar{m}^b) -(2\mu + 4\gamma )K^{(n)}_{ab} \bar{m}^a\bar{m}^b\\
&= -(\pa_\rho K^{(n)}_{ab} -   \theta^{(n)} K^{(n)}_{ab} )\bar{m}^a\bar{m}^b .
\end{aligned}
\end{equation}
Since $\Psi_{N4} = e^{-2i\chi}\Psi_4$ and $e^{-2i\chi}=O(\rho)$, we have that $\mr{\Psi}_{N4} = \mr{\Psi}_4$. However, this is not true for the sub-leading orders. Then, to compute the sub-leading order of $\Psi_{N4}$, we proceed as follows. We know that
\begin{equation}
   \Psi_{N4}= -\pa_\rho\lambda_N - 2\mu_N\lambda_N,
\end{equation}
and since the Lorentz transformations of class II and III do not change the vector $n$, we have that $\lambda_N = K^{(n)}_{ab}\bar{m}^a_N \bar{m}^b_N$. Then
\begin{equation}
\begin{aligned}
\Psi_{N4}^{(1)}&= -2\lambda_N^{(2)} - 2\mu_N\lambda_N^{(1)} -2\mu^{(1)}_N\lambda_N\\
&=2\mr{\mu}(2\mr{\lambda}\mr{\mu} + \mr{\Psi}_4) +2\mr{\lambda}(\mr{\mu}^2 +\mr{\lambda}\mr{\bar{\lambda}}) - 2\lambda^{(2)}_N
\end{aligned}
\end{equation}
where
\begin{equation}
\begin{aligned}
\lambda^{(2)}_N&= {}^{(2)}\mr{K}^{(n)}_{ab} \mr{\bar{E}}^a \mr{\bar{E}}^b - d_{ab} (\lambda \mr{q}^{ab} + \mtn\mr{\bar{E}}^a\mr{\bar{E}}^b) + 2\lambda^2 \bar{\lambda} + 6\mu^2\lambda \\
& +(2\mr{\lambda}\mr{\mu} + \mr{\Psi}_4)\mr{\mu}  +(\mr{\mu}^2 +\mr{\lambda}\mr{\bar{\lambda}})\mr{\lambda}  .
\end{aligned}
\end{equation}
Putting these contributions together, we obtain
\begin{equation}
    \Psi_{N4}^{(1)} = -\Bigl(3d^{(1)}_{ab} -2d_{ab}\mtn +2(\mtn)^2\mr{K}_{ab}^{(n)}\Bigl)\mr{\bar{E}}^a \mr{\bar{E}}^b.
\end{equation}

\subsection*{Radial expansion of $\Psi_k$}
Then, using the Newman-Penrose radial equations and the results obtained above, we have
\begin{equation}
\begin{aligned}
{\Psi}_{N4} &=- \cT_{ab} \ \mr{\bar{E}}^a {\bar{E}}^b +\rho \Psi_{N4}^{(1)} + o(\rho), \\
{\Psi}_{N3} &= - \cP_a\ \mr{\bar{E}} + \rho\Bigl( \eth \mr{\Psi}_4 - \mr{\mu}\mr{\Psi}_3\Bigl) + o(\rho),\\
{\Psi}_{N2} &= -\cA_{ab}\mr{E}^a\mr{\bar{E}}^b + \rho\Bigl(\eth\mr{\Psi}_3 -3\mr{\mu}\mr{\Psi}_2 + \mr{\sigma}\mr{\Psi}_4 \Bigl) + o(\rho), \\
{\Psi}_{N1} &= -\cJ_a\ \mr{E}^a  + \rho \Bigl( \eth\mr{\Psi}_2 -2\mr{\mu}\mr{\Psi}_1 +2\mr{\sigma} \mr{\Psi}_3 \Bigl) + o(\rho), \\
\Psi_{N0} &= -\cN_{ab}\ \mr{E}^a \mr{E}^b + \rho\Bigl( \eth \mr{\Psi}_1 -\mr{\mu}\mr{\Psi}_0+ 3\mr{\sigma}\mr{\Psi}_2 \Bigl) + o(\rho).
\end{aligned}
\end{equation}
The knowledge of the sub-leading terms of the spin coefficients and Weyl scalars allows to compute the spin-$-3$ charge and its evolution equation. The evaluation and the transformation properties of the charges with spin $s\geq3$ will be investigated in a future work. In section \ref{cov_obs}, we computed the boost weights of the covariant functional, obtaining $w(\cT_{ab})=-2,\ w(\cP_{a})=-1,\ w(\cA_{ab})=0,\  w(\cJ_{a})=1,\ w(\cN_{ab})=2$. Since the complex dyad $(\mr{E}^a, \mr{\bar{E}}^a)$ has vanishing boost weights and $(1,-1)$ spin weights and in the transformation in \eqref{weyl_oa} we take $\omega_1=-1$ and $\omega_0=0$, we have the following $(s,w,\omega)$-weights 
\begin{equation}
\begin{aligned}
\Psi_{4}=(-2,-2,-5),& \qquad \Psi_3=(-1,-1,-4), \qquad \Psi_2=(0,0,-3), \\
&\Psi_1=(1,1,-2), \qquad \Psi_0=(2,2,-1),
\end{aligned}
\end{equation}
or compactly $\Psi_k=(2-k,2-k, -k-1)$.

\subsection*{Bianchi identities}\label{ev_BI}
In the previous subsection we derived the leading expressions of the Weyl scalars, and now we are interested in the evolution equation of the latter. Using the Bianchi identities, these yield
\begin{equation}
\begin{aligned}
(\pa_v - 4\mr{\varrho} - 2\mr{\epsilon} ) \mr{\Psi}_1 &= (\bar{\eth}+\mpi)\mr{\Psi}_0,\\
(\pa_v -3\mr{\varrho})\mr{\Psi}_2  &=(\bar{\eth} + 2 \mpi)\mr{\Psi}_1 -\mr{\lambda}\mr{\Psi}_0, \\
(\pa_v -2\mr{\varrho} + 2\mr{\epsilon}) \mr{\Psi}_3 &= (\bar{\eth}+ 3\mpi) \mr{\Psi}_2 -2\mr{\lambda}\mr{\Psi}_1, \\
(\pa_v - \mr{\varrho} + 4\mr{\epsilon}) \mr{\Psi}_4  &= (\bar{\eth} + 4\mpi )\mr{\Psi}_3 -3\mr{\lambda}\mr{\Psi}_2. 
\end{aligned}
\label{Psi_ee}
\end{equation}
These equations can be written in a recursive formula in terms of the weighted operators defined in \cite{Penrose:1985bww},
\begin{equation}
\pth_\scr{C} \mr{\Psi}_n = \bar{\eth}_{\scr{C}} \mr{\Psi}_{n-1} - (n-1)\mr{\lambda}\mr{\Psi}_{n-2}, \qquad \text{with}\ n=1,2,3,4,
\end{equation}
where in our case
\begin{equation}
    \pth_{\scr{C}}  = \pth + (\omega -2w)\mr{\rho}, \qquad\text{and}\qquad \bar{\eth}_{\scr{C}}  = \bar{\eth} - (\omega -w+s)\mpi.
\end{equation}
Notice that the operator $\pth_{\scr{C}}$ ($\pth_{\scr{C}}'$) is a boost-raising (-lowering) operator, since it raises (lowers) the boost weight of $+1$ ($-1$), while $\eth_{\scr{C}}$ (resp. $\bar{\eth}_{\scr{C}}$) is a spin-raising (-lowering) operator, since it raises (lowers) the spin weight of $+1$ ($-1$). Moreover, in our convention $\pth_{\scr{C}}'$ lowers the Weyl weight of $-2$ while  $\eth_{\scr{C}}$ and $\bar{\eth}_{\scr{C}}$ lower the Weyl weight of $-1$. Now, we want to rearrange the evolution Bianchi identities in \eqref{Psi_ee} to recover the evolution equations of the boost-weighted covariant functionals listed in subsection \ref{sub_sec_3.4}. For this purpose, let us explicitly work out the terms of the form
\begin{equation}
    (\pa_v - 2s\mr{\epsilon}_c) \mr{\Psi}_s , \qquad \text{where}\quad \mr{\epsilon}_{r,c}=\f{\mr{\epsilon} \pm \mr{\bar{\epsilon}}}{2}. \label{eps_C}
\end{equation}
In the first Bianchi identity, we have
\begin{equation}
\begin{aligned}
\pa_v\mr{\Psi}_1 -2\mr{\epsilon}_c \mr{\Psi}_1&= -\pa_v(\cJ_a\mr{E}^a) + 2\mr{\epsilon}_c \cJ_a\mr{E}^a\\
&= -\pa_v(\cJ_a)\mr{E}^a + 2\mr{\epsilon}_c \cJ_a\mr{E}^a
- \sigma \cJ_a \mr{\bar{E}}^a - \cJ_a(\mr{\varrho} +2\mr{\epsilon}_c) \mr{E}^a\\
&= -\pa_v(\cJ_a)\mr{E}^a - \mr{\sigma} \cJ_a \mr{\bar{E}}^a - \cJ_a \mr{\varrho} \mr{E}^a .   
\end{aligned}
\end{equation}
The term \eqref{eps_C} in the evolution equation for $\cP_a$ yields
\begin{equation}
\begin{aligned}
\pa_v \mr{\Psi}_3 + 2\mr{\epsilon}_c \mr{\Psi}_3 &= -\pa_v(\cP_a\mr{\bar{E}}^a) - 2\mr{\epsilon}_c \cP_a \mr{\bar{E}}^a\\
&= -\mr{\bar{E}}^a\pa_v\cP_a  -\cP_a(\mr{\bar{\sigma}} \mr{E}^a +(\mr{\varrho} -2\mr{\epsilon}_c)\mr{\bar{E}}^a) - 2\mr{\epsilon}_c \cP_a \mr{\bar{E}}^a\\
&= -\mr{\bar{E}}^a\pa_v\cP_a - \cP_a\mr{\bar{\sigma}} \mr{E}^a -\cP_a\mr{\varrho}\mr{\bar{E}}^a,
\end{aligned}
\end{equation}
and finally, in the last Bianchi identity we have
\begin{equation}
\begin{aligned}
 \pa_v \mr{\Psi}_4 + 4\mr{\epsilon}_c \mr{\Psi}_4 &= -\pa_v(\cT_{ab}\mr{\bar{E}}^a\bar{E}^b) - 4\mr{\epsilon}_c \cT_{ab}\mr{\bar{E}}^a\bar{E}^b\\
&= -(\pa_v\cT_{ab})\mr{\bar{E}}^a\bar{E}^b  -\cT_{ab}\pa_v(\mr{\bar{E}}^a\bar{E}^b) - 4\mr{\epsilon}_c \cT_{ab}\mr{\bar{E}}^a\bar{E}^b\\
&= -(\pa_v\cT_{ab}) \mr{\bar{E}}^a \bar{E}^b  -2\mr{\varrho} \cT_{ab}\mr{\bar{E}}^b\mr{\bar{E}}^a,    
\end{aligned}
\end{equation}
where we used the metric equation
\begin{equation}
\pa_v \mr{E}^a = \mr{\sigma} \mr{\bar{E}}^a + (\mr{\varrho} +\mr{\epsilon}-\mr{\bar{\epsilon}}) \mr{E}^a.
\end{equation}
Thus, substituting the above relations into \eqref{Psi_ee}, using the expressions of the spin coefficients listed in \eqref{list_sc}, the properties of the dual symmetry introduced in subsection \ref{area_dual} and the relations in \eqref{DD}, we finally obtain
\begin{equation}
\begin{aligned}
&\Bigl[\pa_v\cJ_a -\Bigl(\mkell-\f{3}{2}\mthell\Bigl)\cJ_a -\f{1}{2}(\mr{K}^{(\ell)b}_{a} \cJ_b+ \mr{\tilde{K}}^{(\ell)b}_{a} \tilde{\cJ}_b )- (\mr{D}_b -\mpi_b)\cN^{\ b}_a\Bigl]\mr{E}^a=0,\\
&\Bigl[\pa_v\cA_{ab} +\f{1}{2}\mthell\cA_{ab} - \cA_{c(a}(\mr{K}^{(\ell)c}_{b)} -\mr{\wt{K}}^{(\ell)c}_{b)} ) -(\mr{D}_a + \mpi_a)\cJ_b + \mr{\sigma}^{(n)}_{ac}\cN^{c}_{\ b} \Bigl]\mr{E}^a\mr{\bar{E}}^b=0,\\
&\Bigl[\pa_v\cP_a + \mr{\mu}_{(\ell)} \cP_a -\f{1}{2}(\mr{K}^{(\ell)b}_{a}\cP_b + \mr{\tilde{K}}^{(\ell)b}_{a}\tilde{\cP}_b) -(\mr{D}_b +3\mpi_b)\cA_a^{\ b} +2\mr{\sigma}^{(n)}_{ab}\cJ^b\Bigl]\mr{\bar{E}}^a=0,\\
&\Bigl[\pa_v\cT_{ab} +\Bigl(2\mkell -\f{1}{2}\mthell \Bigl) \cT_{ab} - (\mr{D}_{(a}+5\mpi_{(a}) \cP_{b)} +3\mr{\sigma}^{(n)}_{ca}\cA_b^{\ c} \Bigl]\mr{\bar{E}}^a \mr{\bar{E}}^b=0.
\end{aligned}
\label{ev_cf}
\end{equation}
In particular, we have $\mr{\wt{\theta}}^{(\ell)}\wt{\cJ}_a = -\mthell \cJ_a$, $\mr{\wt{\theta}}^{(\ell)}\wt{\cP}_a =- \mthell \cP_a$ and also
\begin{equation}
\mr{\sigma}^{(\ell)}_{ab}\cJ^b = \mr{\wt{\sigma}}^{(\ell)}_{ab}\wt{\cJ}^b, \qquad \text{and}\qquad  \mr{\sigma}^{(\ell)}_{ab}\cP^b = \mr{\wt{\sigma}}^{(\ell)}_{ab}\wt{\cP}^b.
\end{equation}
Therefore, the first and the third equations in \eqref{ev_cf} become
\begin{equation}
\begin{aligned}
    \pa_v \cJ_a -\Bigl(\mkell-\f{3}{2}\mthell\Bigl)\cJ_a - \mr{\sigma}^{(\ell)b}_{a} \cJ_b- (\mr{D}_b -\mpi_b)\cN^{\ b}_a =0
\end{aligned}
\end{equation}
and
\begin{equation}
\begin{aligned}
\pa_v\cP_a + \mr{\mu}_{(\ell)} \cP_a -\mr{\sigma}^{(\ell)b}_{a}\cP_b  -\f{1}{2}(\pa_a +3\mpi_a)\cA - \f{1}{2}(\wt{\pa}_a +3\mr{\wt{\pi}}_a)\wt{\cA} +2\mr{\sigma}^{(n)}_{ab}\cJ^b =0,
\end{aligned}
\end{equation}
respectively. In particular, by noticing that $\cA_{ab}=\f{1}{2}(\cA\mr{q}_{ab}+\wt{\cA} \varepsilon_{ab})$, the second Bianchi identity can be split in two equations: its symmetric part
\begin{equation}
    \pa_v \cA + \f{3}{2}\mthell\cA - (\mr{D}_a + \mpi_a)\cJ^a +\mr{\sigma}^{(n)}_{ab}\cN^{ab}=0,
\end{equation}
and its anti-symmetric part
\begin{equation}
    \pa_v \wt{\cA} + \f{3}{2}\mthell\wt{\cA} - (\mr{D}_a + \mpi_a)\wt{\cJ}^a +\mr{\sigma}^{(n)}_{ab}\wt{\cN}^{ab}=0.
\end{equation}

\section{Useful formulae} \label{formulae}
In this appendix, we provide the derivation of some formulae used in the paper. Let us start by computing the behaviour of the boundary connection under symmetry transformations. One obtains that
\begin{equation}
\begin{aligned}
\var_{(\tau,Y)} \mr{\Gamma}^c_{ab}&= \f{1}{2}\mr{q}^{cd}(\mr{D}_a \var_\xi \mr{q}_{bd} + \mr{D}_b \var_\xi \mr{q}_{ad}-\mr{D}_d \var_\xi \mr{q}_{ab})\\
&= \mr{q}^{cd}\Bigl( \mr{D}_a\mr{D}_{(b}Y_{d)}+ \mr{D}_b \mr{D}_{(a} Y_{d)} -\mr{D}_d \mr{D}_{(a}Y_{b)} + \tau\mr{D}_a \mr{K}^{(\ell)}_{bd} + \tau\mr{D}_b \mr{K}^{(\ell)}_{ad}\\
&\quad -\tau\mr{D}_d \mr{K}^{(\ell)}_{ab}+ \mr{K}^{(\ell)}_{bd} \mr{D}_a \tau + \mr{K}^{(\ell)}_{ad}\mr{D}_b \tau - \mr{K}^{(\ell)}_{ab}\mr{D}_d \tau\Bigl)\\
&= \mr{D}_{(a}\mr{D}_{b)}Y^c + \f{1}{2}(R_{bda}^{\quad\ c} + R_{adb}^{\quad\ c} )Y^d + 2\tau\mr{D}_{(a} \mr{K}^{(\ell)c}_{b)}-\tau\mr{D}^c \mr{K}^{(\ell)}_{ab}\\
&\quad + 2\mr{K}^{(\ell)c}_{(b} \mr{D}_{a)} \tau- \mr{K}^{(\ell)}_{ab}\mr{D}^c \tau,
\end{aligned}
\end{equation}
where we used the following commutators
\begin{equation}
    [\mr{D}_a, \mr{D}_b]Y_c= \mr{R}_{abc}^{\quad\ d}Y_d,\qquad [\mr{D}_a, \mr{D}_b]Y^c= \mr{R}^c_{\ dab}Y^d.
\end{equation}
In particular, one infers that the Christoffel symbols possess the following linear anomalies
\begin{equation}
\Delta_\xi \mr{\Gamma}^c_{ab} = 2\mr{K}^{(\ell)c}_{(b} \mr{D}_{a)} \tau- \mr{K}^{(\ell)}_{ab}\mr{D}^c \tau ,\qquad \Delta_\xi \mr{\Gamma}^b_{ab} = \mthell\mr{D}_{a} \tau.
\label{Gamma_anom}
\end{equation}
Now, having established the behaviour of the Christoffel symbols under symmetry transformations, the Ricci tensor associated with the boundary metric possesses the following anomaly
\begin{equation}
\begin{aligned}
\Delta_\xi\mr{\cR}_{ab} &= \Delta_\xi \Bigl(\pa_c\mr{\Gamma}^c_{ab}-\pa_b\mr{\Gamma}^c_{ac} + \mr{\Gamma}^d_{cd}\mr{\Gamma}^c_{ab} - \mr{\Gamma}^c_{bd}\mr{\Gamma}^d_{ac}\Bigl)\\
&=  2\mr{D}_{(a} \mr{K}^{(\ell)c}_{b)}\pa_c\tau-\mr{D}^c \mr{K}^{(\ell)}_{ab} \pa_c\tau + 2\pa_c(\mr{K}^{(\ell)c}_{(b} \mr{D}_{a)} \tau)- \pa_c(\mr{K}^{(\ell)}_{ab}\mr{D}^c \tau)\\
&\quad -\pa_b\tau\mr{D}_{a} \mthell - \pa_b(\mthell \mr{D}_{a} \tau)+ 2\mr{\Gamma}^d_{cd}\mr{K}^{(\ell)c}_{(b} \mr{D}_{a)} \tau- \mr{\Gamma}^d_{cd} \mr{K}^{(\ell)}_{ab}\mr{D}^c \tau \\
&\quad +\mthell\mr{\Gamma}^c_{ab}\pa_c\tau + \mr{\Gamma}^c_{bd} \mr{K}^{(\ell)}_{ac}\mr{D}^d \tau -2\mr{\Gamma}^d_{ac}\mr{K}^{(\ell)c}_{(b} \mr{D}_{d)} \tau+ \mr{\Gamma}^d_{ac} \mr{K}^{(\ell)}_{db}\mr{D}^c \tau\\
&\quad - 2\mr{\Gamma}^c_{bd}\mr{K}^{(\ell)d}_{(c} \pa_{a)} \tau\\
&=  2\mr{D}_{(a} \mr{K}^{(\ell)c}_{b)}\pa_c\tau-2\mr{D}^c \mr{K}^{(\ell)}_{ab} \pa_c\tau + 2\mr{D}_c(\mr{K}^{(\ell)c}_{(b} \mr{D}_{a)} \tau)- \mr{K}^{(\ell)}_{ab}\mr{D}^c\pa_c \tau\\
& \quad -2\mr{D}_{(a} \mthell\pa_{b)}\tau - \mthell \mr{D}_b\pa_{a} \tau.
\end{aligned}
\end{equation}
Tracing the equation above, the anomaly of the boundary Ricci scalar reads as
\begin{equation}
\begin{aligned}
\Delta_\xi\mr{\cR} &= 4\mr{D}_{a} \mr{K}_{(\ell)}^{ab}\pa_b\tau-4\mr{D}^c \mthell \pa_c\tau + 2\mr{K}_{(\ell)}^{ab} \mr{D}_a \pa_{b} \tau - 2\mthell\mr{D}^c\pa_c \tau.
\end{aligned}
\end{equation}

\bibliography{bib}

\end{document}